\PassOptionsToPackage{unicode}{hyperref}
\PassOptionsToPackage{hyphens}{url}
\documentclass[
]{article}
\usepackage{amsmath,amssymb}
\usepackage{lmodern}
\usepackage{iftex}
\ifPDFTeX
  \usepackage[T1]{fontenc}
  \usepackage[utf8]{inputenc}
  \usepackage{textcomp} 
\else 
  \usepackage{unicode-math}
  \defaultfontfeatures{Scale=MatchLowercase}
  \defaultfontfeatures[\rmfamily]{Ligatures=TeX,Scale=1}
\fi
\IfFileExists{upquote.sty}{\usepackage{upquote}}{}
\IfFileExists{microtype.sty}{
  \usepackage[]{microtype}
  \UseMicrotypeSet[protrusion]{basicmath} 
}{}
\makeatletter
\@ifundefined{KOMAClassName}{
  \IfFileExists{parskip.sty}{%
    \usepackage{parskip}
  }{
    \setlength{\parindent}{0pt}
    \setlength{\parskip}{6pt plus 2pt minus 1pt}}
}{
  \KOMAoptions{parskip=half}}
\makeatother
\usepackage{xcolor}
\usepackage[margin=1.0in]{geometry}
\usepackage{longtable,booktabs,array}
\usepackage{calc} 
\usepackage{etoolbox}
\makeatletter
\patchcmd\longtable{\par}{\if@noskipsec\mbox{}\fi\par}{}{}
\makeatother
\IfFileExists{footnotehyper.sty}{\usepackage{footnotehyper}}{\usepackage{footnote}}
\makesavenoteenv{longtable}
\usepackage{graphicx}
\makeatletter
\def\maxwidth{\ifdim\Gin@nat@width>\linewidth\linewidth\else\Gin@nat@width\fi}
\def\maxheight{\ifdim\Gin@nat@height>\textheight\textheight\else\Gin@nat@height\fi}
\makeatother
\setkeys{Gin}{width=\maxwidth,height=\maxheight,keepaspectratio}
\makeatletter
\def\fps@figure{htbp}
\makeatother
\providecommand{\tightlist}{%
  \setlength{\itemsep}{0pt}\setlength{\parskip}{0pt}}
\newlength{\cslhangindent}
\newlength{\csllabelwidth}
\newlength{\cslentryspacingunit} 
\newenvironment{CSLReferences}[2] 
 {
  \setlength{\parindent}{0pt}
  \ifodd #1
  \let\oldpar\par
  \def\par{\hangindent=\cslhangindent\oldpar}
  \fi
  \setlength{\parskip}{#2\cslentryspacingunit}
 }%
 {}
\usepackage{calc}

\newcommand{\CSLLeftMargin}[1]{\parbox[t]{\csllabelwidth}{#1}}
\newcommand{\CSLRightInline}[1]{\parbox[t]{\linewidth - \csllabelwidth}{#1}\break}

\usepackage{setspace}
\usepackage[left, pagewise]{lineno}
\usepackage[labelfont=bf]{caption}
\usepackage{booktabs}
\usepackage{longtable}
\usepackage{array}
\usepackage{multirow}
\usepackage{wrapfig}
\usepackage{float}
\usepackage{colortbl}
\usepackage{pdflscape}
\usepackage{tabu}
\usepackage{threeparttable}
\usepackage{threeparttablex}
\usepackage[normalem]{ulem}
\usepackage{makecell}
\usepackage{xcolor}
\usepackage{pdfpages} 
\usepackage{pgffor} 

\makeatletter
\AtBeginDocument{\let\LS@rot\@undefined}
\makeatother

\def\supplementfilename{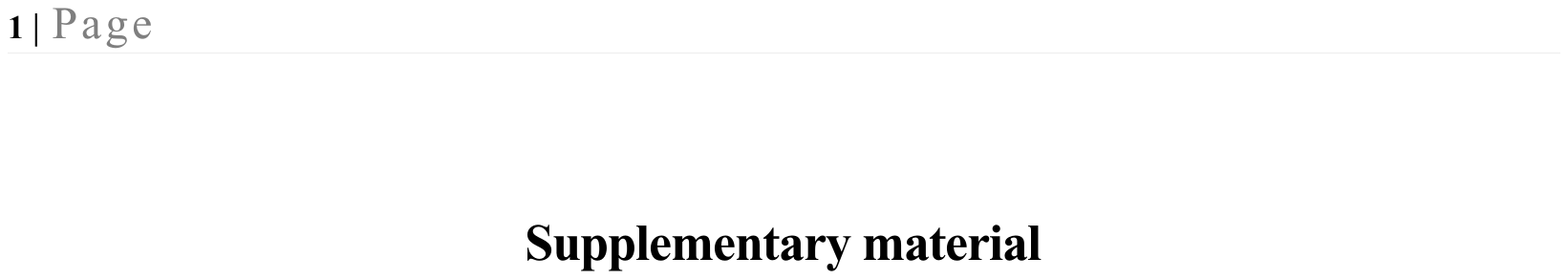}

\pdfximage{\supplementfilename}
\def\numbersupplementpages{\the\pdflastximagepages}

\newif\ifarXiv
\arXivtrue 
\ifLuaTeX
  \usepackage{selnolig}  
\fi
\IfFileExists{bookmark.sty}{\usepackage{bookmark}}{\usepackage{hyperref}}
\IfFileExists{xurl.sty}{\usepackage{xurl}}{} 
\hypersetup{
  pdftitle={A standardized framework for risk-based assessment of treatment effect heterogeneity in observational healthcare databases},
  pdfauthor={Alexandros Rekkas, MSc\{\^{}1\}; David van Klaveren, PhD\^{}\{2,3\},; Patrick B. Ryan, PhD\^{}4; Ewout W. Steyerberg, PhD\^{}\{3,5\}; David M. Kent, MD\^{}2; Peter R. Rijnbeek, PhD\{\^{}1\}},
  hidelinks,
  pdfcreator={LaTeX via pandoc}}

\title{A standardized framework for risk-based assessment of treatment effect heterogeneity in observational healthcare databases}
\author{Alexandros Rekkas, MSc\({^1}\) \and David van Klaveren, PhD\(^{2,3}\), \and Patrick B. Ryan, PhD\(^4\) \and Ewout W. Steyerberg, PhD\(^{3,5}\) \and David M. Kent, MD\(^2\) \and Peter R. Rijnbeek, PhD\({^1}\)}
\date{}

\begin{document}
\maketitle

\thispagestyle{empty}

\(^1\) Department of Medical Informatics, Erasmus University Medical Center,
Rotterdam, Netherlands

\(^2\) Predictive Analytics and Comparative Effectiveness (PACE) Center, Institute
for Clinical Research and Health Policy Studies (ICRHPS), Tufts Medical Center,
Boston, MA, USA

\(^3\) Department of Public Health, Erasmus University Medical Center, Rotterdam,
Netherlands

\(^4\) Janssen Research and Development, 125 Trenton Harbourton Rd,
Titusville,NJ 08560, USA

\(^5\) Department of Biomedical Data Sciences, Leiden University Medical Center,
Leiden, The Netherlands

\vspace{10mm}

\textbf{Corresponding author}
\singlespacing 
Alexandros Rekkas, MSc

Department of Medical Informatics

Erasmus University Medical Center

3000 CA Rotterdam, P.O. Box 2040

Email: \href{mailto:a.rekkas@erasmusmc.nl}{\nolinkurl{a.rekkas@erasmusmc.nl}}
\onehalfspacing

\vspace{10mm}

\textbf{Funding}

This work has been performed in the European Health Data and
Evidence Network (EHDEN) project. This project has received funding from the
Innovative Medicines Initiative 2 Joint Undertaking (JU) under grant agreement
No 80696. The JU receives support from the European Union's Horizon 2020
research and innovation programme and EFPIA.

\newpage
\newpage

\hypertarget{abstract}{%
\section*{Abstract}\label{abstract}}
\addcontentsline{toc}{section}{Abstract}

\singlespacing

Treatment effects are often anticipated to vary across groups of patients with
different baseline risk. The Predictive Approaches to Treatment Effect
Heterogeneity (PATH) statement focused on baseline risk as a robust predictor of
treatment effect and provided guidance on risk-based assessment of treatment
effect heterogeneity in the RCT setting. The aim of this study was to extend
this approach to the observational setting using a standardised scalable
framework. The proposed framework consists of five steps: 1) definition of the
research aim, i.e., the population, the treatment, the comparator and the
outcome(s) of interest; 2) identification of relevant databases; 3) development
of a prediction model for the outcome(s) of interest; 4) estimation of relative
and absolute treatment effect within strata of predicted risk, after adjusting
for observed confounding; 5) presentation of the results. We demonstrate our
framework by evaluating heterogeneity of the effect of angiotensin-converting
enzyme (ACE) inhibitors versus beta blockers on three efficacy and six safety
outcomes across three observational databases. The proposed framework can
supplement any comparative effectiveness study. We provide a publicly available
R software package for applying this framework to any database mapped to the
Observational Medical Outcomes Partnership Common Data Model. In our
demonstration, patients at low risk of acute myocardial infarction received
negligible absolute benefits for all three efficacy outcomes, though they were
more pronounced in the highest risk quarter, especially for hospitalisation with
heart failure. However, failing diagnostics showed evidence of residual
imbalances even after adjustment for observed confounding. Our framework allows
for the evaluation of differential treatment effects across risk strata, which
offers the opportunity to consider the benefit-harm trade-off between
alternative treatments. It is easily applicable and highly informative whenever
treatment effect estimation in observational data is of interest.

\vspace{10mm}

\textbf{Keywords}: observational data, heterogeneity of treatment effect, risk stratification, subgroup analysis
\newpage 

\hypertarget{introduction}{%
\section{INTRODUCTION}\label{introduction}}

Treatment effects can often vary substantially across individual patients,
causing overall effect estimates to be inaccurate for a significant proportion
of the patients at hand\textsuperscript{1,2}. Understanding
heterogeneity of treatment effects (HTE) has been crucial for both personalized
(or precision) medicine and comparative effectiveness research, giving rise to a
wide range of approaches for its discovery, evaluation and application in
clinical practice. A common approach to evaluating HTE in clinical trials is
through subgroup analyses, which are rarely adequately powered and can lead to
false conclusions of absence of HTE or exaggerate its presence\textsuperscript{3,4}. In addition, patients differ with regard to multiple
characteristics simulatneously, resulting in much richer HTE compared to the one
explored with regular on-variable-at-a-time subgroup analyses {[}Kent, BMJ 2018{]}.

Baseline risk is a summary score inherently related to treatment effect that
can represent more closely the variability in patient characteristics\textsuperscript{3,5--8}. For example, an
invasive coronary procedure---in comparison with medical treatment---improves
survival in patients with myocardial infarction at high (predicted) baseline
risk but not in those at low baseline risk\textsuperscript{9}. It has also been shown
that high-risk patients with pre-diabetes benefit substantially more from a
lifestyle modification program than low-risk patients\textsuperscript{10}.

Recently, systematic guidance on the application of risk-based methods for the
assessment of HTE has been developed for RCT data\textsuperscript{11,12}. After
risk-stratifying patients using an existing or an internally derived prediction
model, risk stratum-specific estimates of relative and absolute treatment effect
are evaluated. Several methods for predictive HTE analysis have been adapted for
use in observational data, but risk-based methods are still not readily
available and have been highlighted as an important future research need\textsuperscript{12}.

The Observational Health Data Science and Informatics (OHDSI) collaborative has
established a global network of data partners and researchers that aim to bring
out the value of health data through large-scale analytics by mapping local
databases to the Observational Medical Outcomes Partnership (OMOP) Common Data
Model (CDM)\textsuperscript{13,14}. A standardized framework
applying current best practices for comparative effectiveness studies within the
OHDSI setting has been proposed\textsuperscript{15}. This framework was successfully
implemented on a large scale for estimation of average effects of all first-line
hypertension treatment classes on a total of 52 outcomes of interest across a
global network of nine observational databases\textsuperscript{16}.

We aimed to develop a framework for risk-based assessment of treatment effect
heterogeneity in observational healthcare databases, extending the existing
methodology developed for the RCT setting. We implemented the framework in a
publicly available package providing an out-of-the-box solution for implementing
such analyses at scale within any observational database mapped to OMOP-CDM. In
a case study we analyzed heterogeneity of the effects of first-line hypertension
treatment.

\hypertarget{results}{%
\section{RESULTS}\label{results}}

\begin{figure}
\includegraphics[width=1\linewidth]{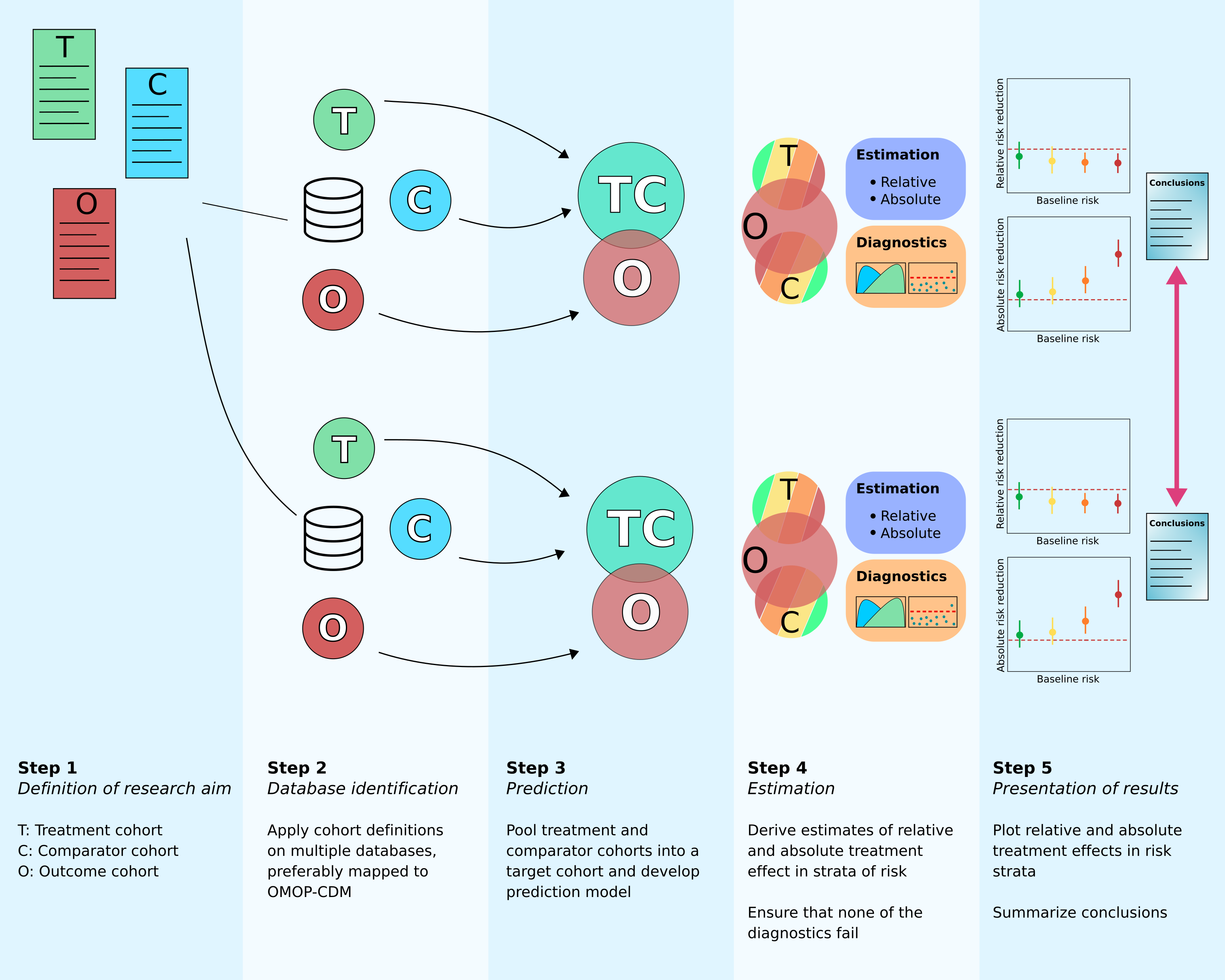} \caption{\textbf{Framework execution diagram}. Illustration of how the framework is applied on two observational databases, preferably mapped to OMOP-CDM.}\label{fig:graphicalAbstract}
\end{figure}

The proposed framework defines 5 distinct steps: 1) definition of the research
aim; 2) identification of the databases within which the analyses will be
performed; 3) prediction of outcomes of interest; 4) estimation of absolute and
relative treatment effects within risk strata; 5) presentation of the
results. We developed an open-source R-package for the implementation of the
proposed framework and made it publicly available
(\url{https://github.com/OHDSI/RiskStratifiedEstimation}). An overview of the entire
framework can be found in Figure \ref{fig:graphicalAbstract}.

As a demonstration, we evaluated if our proposed method was able to identify
treatment effect heterogeneity of ACE inhibitors compared to beta blockers using
acute myocardial infarction (MI) risk quarter specific effect estimates, both on
the relative and on the absolute scale. We focused on three efficacy outcomes
(acute MI, hospitalization with heart failure and ischemic or hemorrhagic
stroke) and six safety outcomes (hypokalemia, hyperkalemia, hypotension,
angioedema, cough and abnormal weight gain). We used data from three US-based
claims databases.

\hypertarget{step-1-general-definition-of-the-research-aim}{%
\subsection{Step 1: General definition of the research aim}\label{step-1-general-definition-of-the-research-aim}}

We considered the following research aim: ``compare the effect of ACE inhibitors
(\(T\)) to the effect of beta blockers (\(C\)) in patients with established hypertension
with respect to nine outcomes (\(O_1,\dots,O_9\))''. The cohorts are:

\begin{itemize}
\tightlist
\item
  Treatment cohort: Patients receiving any drug within the ACE inhibitor class
  with at least one year of follow-up before treatment initiation and a recorded
  hypertension diagnosis within that year.
\item
  Comparator cohort: Patients receiving any drug within the beta blocker class
  with at least one year of follow-up before treatment initiation and a recorded
  hypertension diagnosis within that year.
\item
  Outcome cohorts: We considered three efficacy and six safety outcome cohorts. These
  were patients in the database with a diagnosis of: acute MI; hospitalization
  with heart failure; ischemic or hemorrhagic stroke (efficacy outcomes);
  hypokalemia; hyperkalemia; hypotension; angioedema; cough; abnormal weight
  gain (safety outcomes). Among the safety outcomes we focus on angioedema and
  cough which are two known adverse events linked to treatment with ACE
  inhibitors\textsuperscript{17}. Results on the rest of the safety outcomes
  are included in the supplement.
\end{itemize}

All cohort definitions were identical to the ones used in the multinational
study that provided overall treatment effect estimates comparing all
anti-hypertensive drug classes with each other\textsuperscript{16}. More information
can be found in the supplementary material.

\hypertarget{step-2-identification-of-the-databases}{%
\subsection{Step 2: Identification of the databases}\label{step-2-identification-of-the-databases}}

For our demonstration we used data from three US claims databases, namely IBM
MarketScan Commercial Claims and Encounters (CCAE), IBM MarketScan Medicaid
(MDCD), and IBM MarketScan Medicare Supplemental Beneficiaries (MDCR). Our
analyses included a total of
924,459,
107,046, and
106,905
patients initiating treatment with ACE inhibitors and
465,763,
76,546, and
73,213
patients initiating treatment with beta blockers in CCAE, MDCD and MDCR
respectively (Table \ref{tab:incidenceOverall}). Adequate numbers of patients
were included in all strata of predicted acute MI risk (Supplementary Tables
S1-S3).

\begingroup\fontsize{7}{9}\selectfont

\begin{longtable}[t]{llrrrrr}
\caption{\label{tab:incidenceOverall}Number of patients, person years and events within quarters of predicted risk for acute MI for the three efficacy outcomes of the study (acute MI, hospitalization with heart failure and ischemic or hemorrhagic stroke) across the three databases.}\\
\toprule
\multicolumn{1}{c}{ } & \multicolumn{3}{c}{ACE inhibitors} & \multicolumn{3}{c}{Beta blockers} \\
\cmidrule(l{3pt}r{3pt}){2-4} \cmidrule(l{3pt}r{3pt}){5-7}
Outcome & Patients & Person years & Outcomes & Patients & Person years & Outcomes\\
\midrule
\addlinespace[0.3em]
\multicolumn{7}{l}{\textbf{CCAE}}\\
\hspace{1em}acute myocardial infarction & 924,196 & 1,327,973 & 4,102 & 457,375 & 648,612 & 2,492\\
\hspace{1em}hospitalization with heart failure & 924,459 & 1,328,430 & 3,764 & 465,763 & 660,580 & 3,711\\
\hspace{1em}stroke & 917,501 & 1,319,236 & 3,741 & 464,989 & 659,472 & 2,454\\
\addlinespace[0.3em]
\multicolumn{7}{l}{\textbf{MDCD}}\\
\hspace{1em}acute myocardial infarction & 107,046 & 162,590 & 1,448 & 76,307 & 112,767 & 1,361\\
\hspace{1em}hospitalization with heart failure & 105,544 & 160,237 & 2,819 & 74,649 & 110,455 & 3,005\\
\hspace{1em}stroke & 104,953 & 159,344 & 1,799 & 76,546 & 113,048 & 1,623\\
\addlinespace[0.3em]
\multicolumn{7}{l}{\textbf{MDCR}}\\
\hspace{1em}acute myocardial infarction & 106,905 & 163,260 & 1,764 & 72,733 & 110,821 & 1,480\\
\hspace{1em}hospitalization with heart failure & 106,191 & 162,258 & 3,004 & 73,182 & 111,710 & 3,592\\
\hspace{1em}stroke & 103,531 & 158,369 & 2,323 & 73,213 & 111,613 & 2,241\\
\bottomrule
\end{longtable}
\endgroup{}

\hypertarget{step-3-prediction}{%
\subsection{Step 3: Prediction}\label{step-3-prediction}}

We internally developed separate prediction models for acute MI in all three
databases. The prediction models were estimated on the propensity score matched
(1:1) subset of the sample, using caliper of
0.2
and after excluding patients having the outcome any time prior to treatment
initiation. We chose a 2-year time at risk for patients and developed the
prediction models using LASSO logistic regression with 3-fold cross validation
for hyper-parameter selection.

The models had moderate discriminative performance (internally validated) with
no major issues of overfitting to any cohort except for the case of CCAE, where
the derived prediction model performed better in the comparator cohort (Table
\ref{tab:predictionPerformance}). We also observed lower performance of the
prediction model developed in MDCR compared to the other 2 databases.

\begingroup\fontsize{7}{9}\selectfont

\begin{longtable}[t]{lrrr}
\caption{\label{tab:predictionPerformance}Discriminative ability (c-statistic) of the derived prediction models for acute myocardial infarction in the matched set (development set), the treatment cohort, the comparator cohort and the entire population in CCAE, MDCD and MDCR.}\\
\toprule
Population & CCAE & MDCD & MDCR\\
\midrule
Matched & 0.73 (0.72, 0.74) & 0.78 (0.77, 0.79) & 0.66 (0.65, 0.68)\\
Treatment & 0.69 (0.68, 0.70) & 0.76 (0.75, 0.77) & 0.65 (0.63, 0.66)\\
Comparator & 0.77 (0.76, 0.78) & 0.82 (0.81, 0.82) & 0.64 (0.63, 0.66)\\
Entire population & 0.72 (0.71, 0.73) & 0.79 (0.78, 0.80) & 0.65 (0.64, 0.66)\\
\bottomrule
\end{longtable}
\endgroup{}

\hypertarget{step-4-estimation}{%
\subsection{Step 4: Estimation}\label{step-4-estimation}}

We used patient-level predictions to stratify the sample into four acute MI risk
quarters. Within risk quarters, relative effects were estimated using Cox
regression and absolute effects were derived from the Kaplan-Meier estimate
differences at two years after treatment initiation. To adjust for observed
confounding within each risk quarter, we estimated propensity scores using the
same approach as step 3 and stratified patients into five propensity score
strata. The risk quarter-specific effect estimates were derived by averaging
over the estimates within the propensity score fifths.

\begin{figure}
\includegraphics[width=1\linewidth]{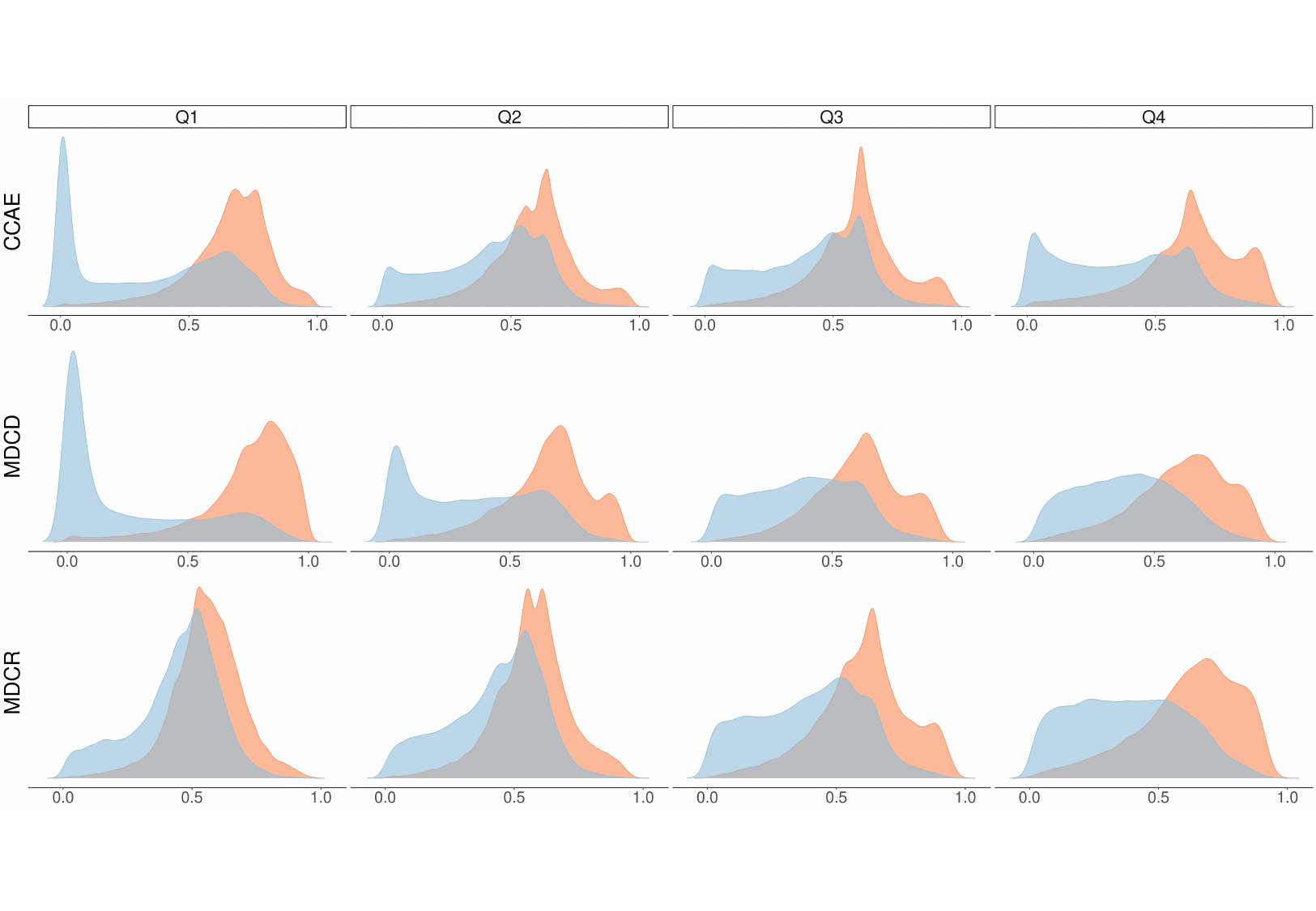} \caption{\textbf{Preference score distributions in quarters of predicted acute MI risk}. The preference score is a transformation of the propensity score that adjusts for prevalence differences between populations. Higher overlap of the preference score distributions indicates that patients in the target and the comparator cohorts are more similar in terms of the predicted probability of receiving treatment (ACE inhibitors).}\label{fig:psDensity}
\end{figure}

\begin{figure}
\includegraphics[width=1\linewidth]{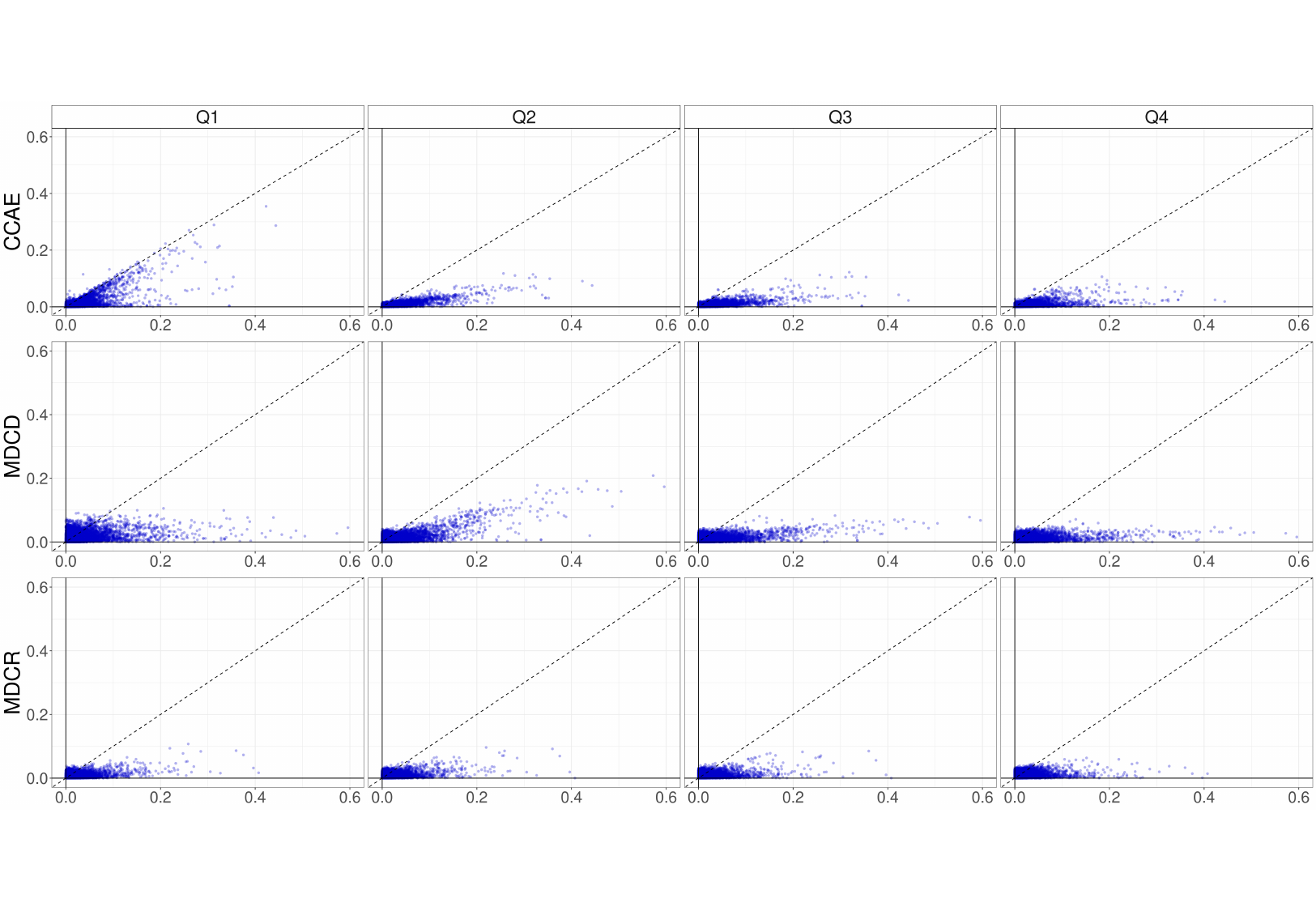} \caption{\textbf{Patient characteristic balance for ACE inhibitors and beta blockers before and after stratification on the propensity scores}. Each dot represents the standardized difference of means for a single covariate before (x-axis) and after (y-axis) stratification. A commonly used rule of thumb suggests that standardized mean differences above 0.1 indicate insufficient covariate balance post propensity score adjustment.}\label{fig:covariateBalance}
\end{figure}

In the lowest acute MI risk quarter of CCAE and MDCD we observed strong
separation of the propensity score distributions, therefore, effect estimates
derived in these strata are not well-supported (Figure
\ref{fig:psDensity}). This problematic behavior is also visible in the covariate
balance plots comparing standardized mean differences of patient characteristics
before and afrer PS adjustment, where in many cases the commonly accepted bound
of
0.1
is violated (Figure \ref{fig:covariateBalance}). This is more pronounced in the
lowest acute MI risk quarter of CCAE, but remains an issue for a small number of
covariates in all CCAE risk strata. This diagnostic also fails for the two lower
acute MI risk quarters of MDCD. Often the persisting imbalances were linked to
pregnancy outcomes, which can be explained by the contraindication of ACE
inhibitors in this condition. Analyses in MDCR passed all diagnostics.

\begin{figure}
\includegraphics[width=1\linewidth]{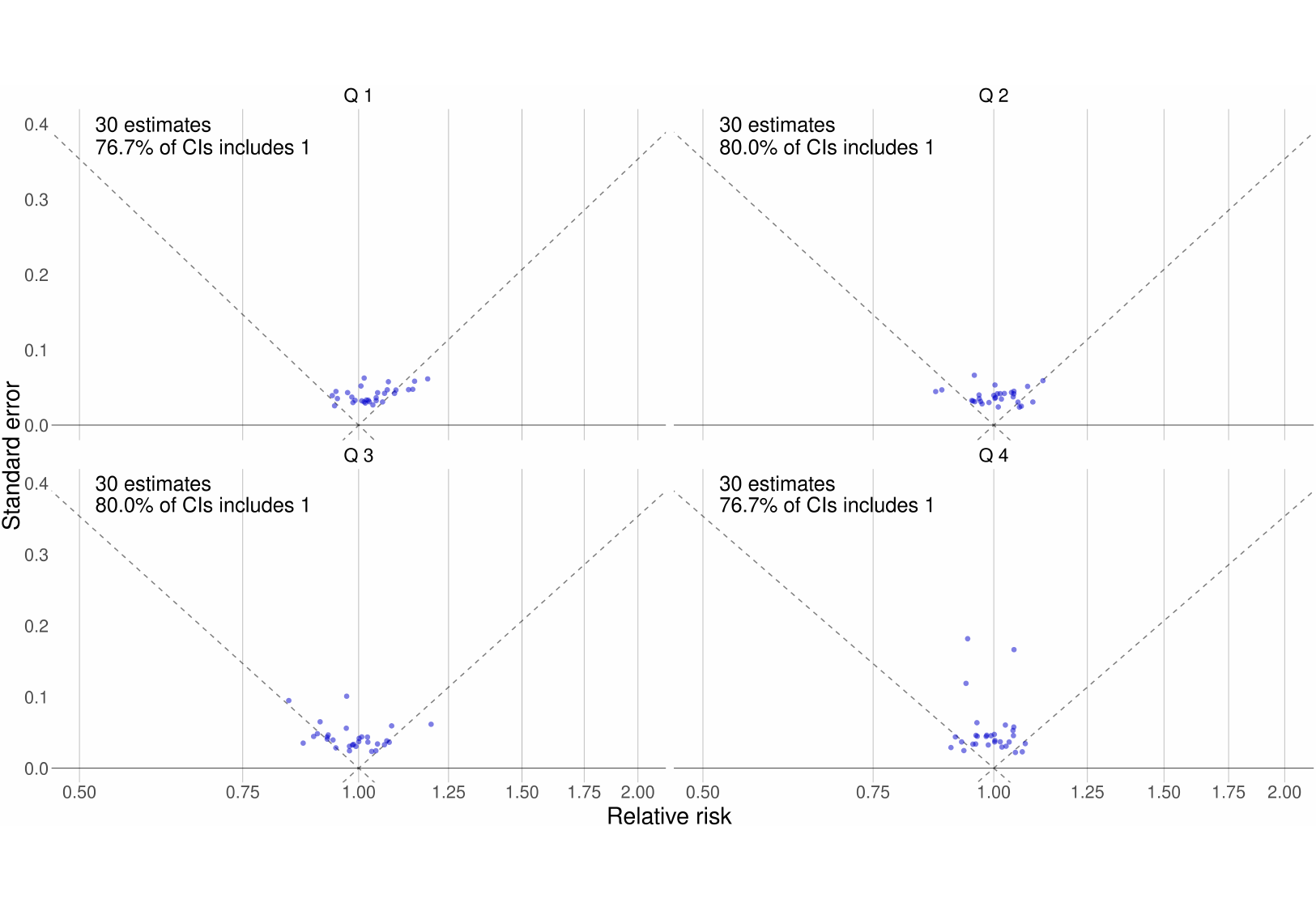} \caption{\textbf{Systematic error}. Effect size estimates for the negative controls (true hazard ratio = 1). Estimates below the diagonal dashed lines are statistically significant (alpha = 0.05) different from the true effect size. A well-calibrated estimator should have the true effect size within the 95 percent confidence interval 95 percent of times.}\label{fig:negativeControls}
\end{figure}

Finally, the distribution of the estimated relative risks with regard to 30
negative control outcomes indicated unresolved confounding within the lowest
acute MI risk quarter of CCAE (Figrue \ref{fig:negativeControls}). Hazard ratios
significantly different than 1 (true effect size) were concentrated in the lower
right part of Figure \ref{fig:negativeControls}: panel Q1. This suggests
significant negative effects of ACE inhibitors compared to beta blockers on
causally unrelated outcomes, pointing at unresolved differences between the two
treatment arms. This was not the case in the other risk quarters of CCAE, or in
any risk quarter of MDCD and MDCR (Supplementary Figures S1-S2).

\hypertarget{step-5-presentation-of-results}{%
\subsection{Step 5: Presentation of results}\label{step-5-presentation-of-results}}

\begin{figure}
\includegraphics[width=1\linewidth]{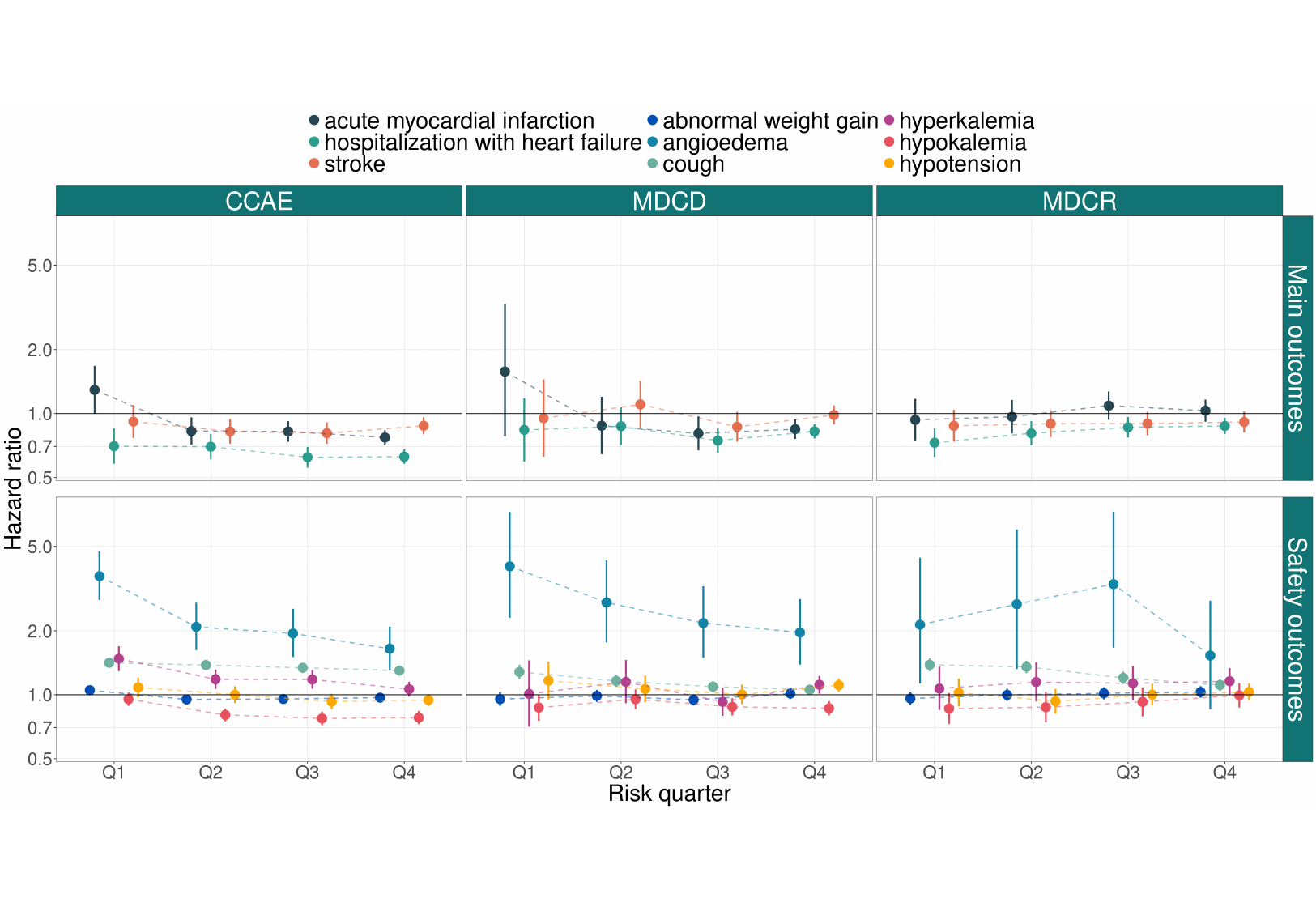} \caption{\textbf{Overview of heterogeneity of ACE inhibitors treatment on the relative scale (hazard ratios) within strata of predicted risk of acute MI}. Values below 1 favor ACE inhibitors, while values above 1 favor beta blockers.}\label{fig:relative}
\end{figure}

The overall estimated hazard ratios for the main outcomes are presented in
Table \ref{tab:overallTable}. For hospitalization with acute MI there was an increasing trend in
favor ACE inhibitors compared to beta blockers on the relative scale (hazard
ratios decreased) with increasing acute MI risk. More specifically, hazard
ratios decreased from
1.29
(1.00 to
1.68;
95\% CI) and
1.58
(0.78 to
3.28;
95\% CI) to
0.77
(0.71 to
0.83;
95\% CI),
0.84
(0.76 to
0.94;
95\% CI) in CCAE and MDCD respectively (Figure \ref{fig:relative}). In MDCR hazard ratios
increased from
0.93
(0.75 to
1.17;
95\% CI) in the lowest MI risk quarter to
1.03
(0.92 to
1.16;
95\% CI). Relative treatment effect estimates for hospitalization with heart
failure favored ACE inhibitors across all risk strata in all databases. In the
case of stroke in CCAE we found quite constant hazard ratios which became weaker
in the highest risk quarter patients
(0.88
with 95\% CI from
0.80
to
0.96).
In the other two databases no significant relative treatment effects were
observed for stroke. In terms of the safety outcomes, we found an increased ACE inhibitor
risk of cough and angioedema on the relative scale across all risk strata. In
the case of cough, this effect decreased with increasing risk of acute MI---from
1.41
(1.37 to
1.46;
95\% CI),
1.28
(1.18 to
1.38;
95\% CI), and
1.38
(1.29 to
1.48;
95\% CI) to
1.30
(1.26 to
1.34;
95\% CI),
1.06
(1.00 to
1.12;
95\% CI), and
1.11
(1.04 to
1.18;
95\% CI) in CCAE, MDCD, and MDCR, respectively.

\begin{table}[H]

\caption{\label{tab:overallTable}Overall hazard ratios.}
\centering
\begin{tabular}[t]{lrrr}
\toprule
Outcome & CCAE & MDCD & MDCR\\
\midrule
acute myocardial infarction & 0.83 (0.79, 0.88) & 0.87 (0.80, 0.96) & 1.02 (0.94, 1.10)\\
hospitalization with heart failure & 0.66 (0.62, 0.69) & 0.86 (0.81, 0.92) & 0.85 (0.80, 0.90)\\
stroke & 0.88 (0.83, 0.93) & 0.97 (0.90, 1.05) & 0.91 (0.85, 0.97)\\
\bottomrule
\end{tabular}
\end{table}

\begin{figure}
\includegraphics[width=1\linewidth]{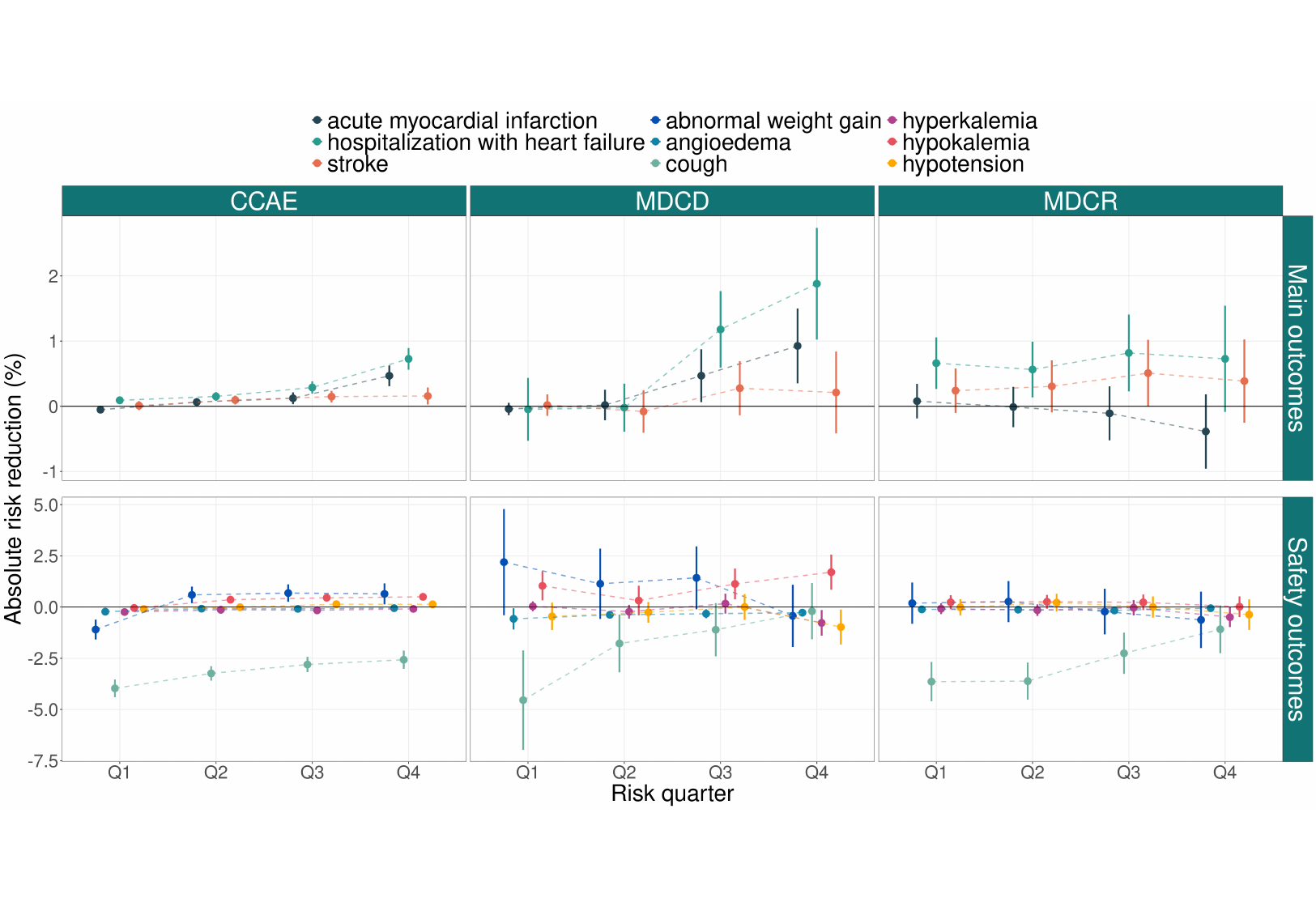} \caption{\textbf{Overview of heterogeneity of ACE inhibitors treatment on the absolute scale within strata of predicted risk of acute MI}. Estimates of absolute treatment effect are derived as the difference in Kaplan-Meier estimates at two years after inclusion. Values above 0 favor ACE inhibitors, while values below 0 favor beta blockers.  }\label{fig:absolute}
\end{figure}

We observed an increasing trend of treatment effect on the absolute scale with
increasing acute MI risk in favor of ACE inhibitors in terms of acute MI in all
databases except for MDCR---from
-0.05\%
(-0.10\% to
-0.005\%;
95\% CI),
-0.04\%
(-0.14\% to
0.05\%;
95\% CI), and
0.08\%
(-0.19\% to
0.34\%;
95\% CI) in the lowest acute MI risk quarter to
0.47\%
(0.31\% to
0.63\%;
95\% CI),
0.93\%
(0.35\% to
1.50\%;
95\% CI), and
-0.39\%
(-0.96\% to
0.18\%;
95\% CI) in the highest acute MI risk quarter in CCAE, MDCD, and MDCR,
respectively (Figure \ref{fig:absolute}). We found no difference on the absolute
scale for stroke across risk strata. Absolute risk differences did not favor ACE
inhibitors compared to beta blockers in terms of cough, even though this effect
again diminished with increasing acute MI risk---from
-3.97\%
(-4.40\% to
-3.54\%;
95\% CI),
-4.54\%
(-6.97\% to
-2.12\%;
95\% CI), and
-3.64\%
(-4.60\% to
-2.68\%;
95\% CI) in the lowest acute MI risk quarter to
-2.57\%
(-3.02\% to
-2.13\%;
95\% CI),
-0.20\%
(-1.58\% to
1.17\%;
95\% CI), and
-1.08\%
(-2.25\% to
0.08\%;
95\% CI) in the highest acute MI risk quarter in CCAE, MDCD, and MDCR,
respectively. In terms of angioedema absolute risk differences were very small
due to the rarity of the outcome.

The results of all the analyses performed can be accessed and assessed through a
publicly available web application (\url{https://data.ohdsi.org/AceBeta9Outcomes}).

\hypertarget{interpretation}{%
\subsection{Interpretation}\label{interpretation}}

The overall benefits of ACE inhibitors compared to beta blockers for acute MI
and hospitalization with heart failure are driven mainly by the higher acute MI
risk patients in CCAE and MDCD, hence the observed increasing patterns of the
absolute treatment benefits. In MDCR we found no significant overall difference
on the relative scale for acute MI and, cosequently, no differences in acute MI
risk strata were observed on any scale. For heart failure, MDCR patients at the
lower half of acute MI risk had lower absolute benefits compared to the patients
at the upper half. Finally, the small overall relative effect for stroke
resulted in smaller absolute benefits of ACE inhibitors across acute MI risk
strata in all databases.

For patients at lower acute MI risk, the cough and angioedema risk increase
related to treatment with ACE inhibitors may be important factors to consider
for medical decision making, given the small benefits observed for the main
outcomes. However, diagnostics failed in lower risk patients within CCAE and
MDCR which renders these conclusions less dependable.

Note that any conclusions drawn are for demonstration purposes only and should
be interpreted under this very limited setting.

\hypertarget{discussion}{%
\section{DISCUSSION}\label{discussion}}

The major contribution of our work is the development of a risk-based framework
for the assessment of treatment effect heterogeneity in large observational
databases. This fills a gap identified in the literature after the development
of guidelines for performing such analyses in the RCT setting\textsuperscript{11,12}. As an additional contribution we developed the software
for implementing this framework in practice and made it publicly available. We
made our software compatible to databases mapped to OMOP-CDM which allows
researchers to easily implement our framework in a global network of healthcare
databases. In our case study we demonstrated the use of our framework for the
evaluation of treatment effect heterogeneity ACE inhibitors compared to beta
blockers on three efficacy and six safety outcomes. We propose that this framework is
implemented any time treatment effect estimation in high-dimensional
observational data is undertaken.

In recent years several methods for the analysis of treatment effect
heterogeneity have been developed in the RCT setting\textsuperscript{18}. However,
low power and restricted prior knowledge on the mechanisms of variation in
treatment effect are often inherent in RCTs, which are usually adequately
powered only for the analysis of the primary outcome. Observational databases
contain a large amount of information on treatment assignment and outcomes of
interest, while also capturing key patient characteristics. They contain readily
available data on patient subpopulations of interest for which no RCT has
focused before either due to logistical or ethical reasons. However,
observational databases can be susceptible to biases, poorly measured outcomes
and missingness, which may obscure true HTE or falsely introduce it when there
is none\textsuperscript{19}. Therefore, inferences on both overall treatment effect
estimates and HTE need to rely on strong, often unverifiable, assumptions,
despite the advancements and guidance on best practices. However, well-designed
observational studies on average replicate RCT results, even though often
differences in magnitude may occur\textsuperscript{24}. Our
framework is in line with the recently suggested paradigm of high-throughput
observational studies using consistent and standardized methods for improving
reproducibility in observational research\textsuperscript{25}.

Our framework highlights the scale dependency of HTE and how it relates to
baseline risk. Treatment effect is mathematically determined by baseline
risk, if we assume a constant non-zero effect size\textsuperscript{26}. Patients
with low baseline risk can only experience minimal benefits, before their risk
is reduced to zero. In contrast, high risk patients are capable of displaying
much higher absolute benefits. This becomes evident when evaluating the effects
of ACE inhibitors on cough and angioedema, compared to treatment with beta
blockers. Despite the small relative cough risk increase of ACE inhibitors, the
large baseline cough risk resulted in larger absolute risk differences, compared
to the other considered outcomes. Conversely, in the case of angioedema, the
substantial relative risk increase with ACE inhibitors only translated in a
small absolute risk increase due to the quite low baseline angioedema risk.

The application of our framework in the case study is for demonstration purposes
and there are several limitations to its conclusions. First, death could be a
competing risk. We could expand our framework in the future to potentially
support subdistribution hazard ratios and cumulative incidence
reductions. Second, we only used the databases readily available to us and not
all the available databases mapped to OMOP-CDM. Therefore, the generalizability
of our results still needs to be explored in future studies. These studies
should also address the particular aspects of the databases at hand, such as
their sampling frame, the completeness of the data they capture and many others
that were not assessed in our demonstration. Third, we did not to correct for
multiplicity when presenting the results. We are interested in presenting trends
in the data and not detecting the specific subgroups within which a non-null
treatment effect is detected. The implementation of our framework, however,
generates all the relevant output required for a researcher to correct for
multiple testing, if that is required.

In conclusion, the case study demonstrates the feasibility of our framework for
risk-based assessment of treatment effect heterogeneity in large observational
data. It is easily applicable and highly informative whenever treatment effect
estimation in high-dimensional observational data is of interest.

\hypertarget{methods}{%
\section{METHODS}\label{methods}}

\hypertarget{step-1-general-definition-of-the-research-aim-1}{%
\subsection{Step 1: General definition of the research aim}\label{step-1-general-definition-of-the-research-aim-1}}

The typical research aim is: ``to compare the effect of treatment to a comparator
treatment in patients with disease with respect to outcomes \(O_1,\dots,O_n\)''.

We use a comparative cohort design. This means that at least three cohorts of
patients need to be defined at this stage of the framework:

\begin{itemize}
\tightlist
\item
  A single treatment cohort (\(T\)) which includes patients with disease receiving
  the target treatment of interest.
\item
  A single comparator cohort (\(C\)) which includes patients with disease receiving
  the comparator treatment.
\item
  One or more outcome cohorts (\(O_1,\dots,O_n\)) that contain patients
  developing the outcomes of interest
\end{itemize}

\hypertarget{methods2}{%
\subsection{Step 2: Identification of the databases}\label{methods2}}

Including in our analyses multiple databases representing the population of
interest potentially increases the generalizability of results. Furthermore, the
cohorts should preferably have adequate sample size with adequate follow-up time
to ensure precise effect estimation, even within smaller risk strata. Other
relevant issues such as the depth of data capture (the precision at which
measurements, lab tests, conditions are recorded) and the reliability of data
entry should also be considered.

\hypertarget{methods3}{%
\subsection{Step 3: Prediction}\label{methods3}}

Our method relies on adequately separating patients into subgroups based on
their baseline risk for the outcomes of interest. Therefore, a model---either an
existing external model adequately validated on an internally developed
one---assigning patient-level risk is required. For internally developing a risk
prediction model we adopt the standardized framework focused on observational
data that ensures adherence to existing guidelines\textsuperscript{27--29}.

We first need to define a target cohort of patients, i.e.~the set of patients on
whom the prediction model will be developed. In our case, the target cohort is
generated by pooling the already defined treatment and comparator cohorts. We
develop the prediction model on the propensity score-matched (1:1) subset of the
pooled sample to avoid differentially fitting between treatment arms, thus
introducing spurious interactions with treatment\textsuperscript{30,31}. We also need to define a set of patients that
experience the outcome of interest, i.e.~the outcome cohort. Finally, we need to
decide the time frame within which the predictions will be carried out, i.e.~the
patients' time at risk. Subsequently, we can develop the prediction model.

It is important that the prediction models display good discriminative ability
to ensure that risk-based subgroups are accurately defined. A performance
overview of the derived prediction models including discrimination and
calibration both in the propensity score matched subset, the entire sample and
separately for treated and comparator patients should also be reported.

\hypertarget{methods4}{%
\subsection{Step 4: Estimation}\label{methods4}}

We estimate treatment effects (both on the relative and the absolute scale)
within risk strata defined using the prediction model of step 3. We often
consider four risk strata, but fewer or more strata can be considered depending on
the available power for accurately estimating stratum-specific treatment
effects. Effect estimation may be focused on the difference in outcomes for a
randomly selected person from the risk stratum (average treatment effect) or for
a randomly selected person from the treatment cohort within the risk stratum
receiving the treatment under study (average treatment effect on the treated).

Any appropriate method for the analysis of relative and absolute treatment
effects can be considered, as long as the this is done consistently in all risk
strata. Common statistical metrics are odds ratios or hazard ratios for relative
scale estimates and differences in observed proportions or differences in
Kaplan-Meier estimates for absolute scale estimates, depending on the problem at
hand. We estimate propensity scores within risk strata which we then use to
match patients from different treatment cohorts or to stratify them into groups
with similar propensity scores or to weigh each patient's contribution to the
estimation process\textsuperscript{32}.

Prior to analyzing results, it is crucial to ensure that all diagnostics are passed
in all risk strata. The standard diagnostics we carry out include analysis of
the overlap of propensity score distributions and calculation of standardized
mean differences of the covariates before and after propensity score
adjustment. Finally, we use effect estimates for a large set of negative control
outcomes (i.e.~outcomes known to not be related with any of the exposures under
study) to evaluate the presence of residual confounding not accounted for by
propensity score adjustment\textsuperscript{25,33,34}.

\hypertarget{methods5}{%
\subsection{Step 5: Presentation of results}\label{methods5}}

In the presence of a positive treatment effect and a well-discriminating
prediction model we expect an increasing pattern of the differences in the
absolute scale, even if treatment effects remain constant on the relative scale
across risk strata. Due to this scale-dependence of treatment effect
heterogeneity, results should be assessed both on the relative and the absolute
scale.

\hypertarget{data-availability}{%
\section{DATA AVAILABILITY}\label{data-availability}}

The claims data are proprietary and are not publicly accessible due to
restricted user agreement. Database descriptions are available in the
Supplementary Material.

\hypertarget{code-availability}{%
\section{CODE AVAILABILITY}\label{code-availability}}

The source code for the R-package that implements our framework can be found at
\url{https://github.com/OHDSI/RiskStratifiedEstimation}. The code for implementing the
proof of concept study presented here is publicly available at
\url{https://github.com/mi-erasmusmc/AceBeta9Outcomes}.

\newpage

\hypertarget{references}{%
\section{REFERENCES}\label{references}}

\setlength{\parindent}{-0.25in}
\setlength{\leftskip}{0.25in}

\noindent

\hypertarget{refs}{}
\begin{CSLReferences}{0}{0}
\leavevmode\vadjust pre{\hypertarget{ref-Rothwell1995}{}}%
\CSLLeftMargin{1 }%
\CSLRightInline{Rothwell PM. \href{https://doi.org/10.1016/s0140-6736(95)90120-5}{Can overall results of clinical trials be applied to all patients?} \emph{The Lancet} 1995; \textbf{345}: 1616--9.}

\leavevmode\vadjust pre{\hypertarget{ref-KRAVITZ2004}{}}%
\CSLLeftMargin{2 }%
\CSLRightInline{Kravitz RL, Duan N, Braslow J. \href{https://doi.org/10.1111/j.0887-378x.2004.00327.x}{Evidence-based medicine, heterogeneity of treatment effects, and the trouble with averages}. \emph{The Milbank Quarterly} 2004; \textbf{82}: 661--87.}

\leavevmode\vadjust pre{\hypertarget{ref-Hayward2006}{}}%
\CSLLeftMargin{3 }%
\CSLRightInline{Hayward RA, Kent DM, Vijan S, Hofer TP. Multivariable risk prediction can greatly enhance the statistical power of clinical trial subgroup analysis. \emph{{BMC} Medical Research Methodology} 2006; \textbf{6}. DOI:\href{https://doi.org/10.1186/1471-2288-6-18}{10.1186/1471-2288-6-18}.}

\leavevmode\vadjust pre{\hypertarget{ref-Kent2018}{}}%
\CSLLeftMargin{4 }%
\CSLRightInline{Kent DM, Steyerberg E, Klaveren D van. \href{https://doi.org/10.1136/bmj.k4245}{Personalized evidence based medicine: Predictive approaches to heterogeneous treatment effects}. \emph{{Bmj}} 2018; : k4245.}

\leavevmode\vadjust pre{\hypertarget{ref-Rothwell2005}{}}%
\CSLLeftMargin{5 }%
\CSLRightInline{ROTHWELL P, MEHTA Z, HOWARD S, GUTNIKOV S, WARLOW C. \href{https://doi.org/10.1016/s0140-6736(05)70156-2}{From subgroups to individuals: General principles and the example of carotid endarterectomy}. \emph{The Lancet} 2005; \textbf{365}: 256--65.}

\leavevmode\vadjust pre{\hypertarget{ref-Kent2007}{}}%
\CSLLeftMargin{6 }%
\CSLRightInline{Kent DM, Hayward RA. \href{https://doi.org/10.1001/jama.298.10.1209}{Limitations of applying summary results of clinical trials to individual patients}. \emph{{JAMA}} 2007; \textbf{298}: 1209.}

\leavevmode\vadjust pre{\hypertarget{ref-Kent2008}{}}%
\CSLLeftMargin{7 }%
\CSLRightInline{Kent DM, Alsheikh-Ali A, Hayward RA. Competing risk and heterogeneity of treatment effect in clinical trials. \emph{Trials} 2008; \textbf{9}. DOI:\href{https://doi.org/10.1186/1745-6215-9-30}{10.1186/1745-6215-9-30}.}

\leavevmode\vadjust pre{\hypertarget{ref-Kent2010}{}}%
\CSLLeftMargin{8 }%
\CSLRightInline{Kent DM, Rothwell PM, Ioannidis JP, Altman DG, Hayward RA. Assessing and reporting heterogeneity in treatment effects in clinical trials: A proposal. \emph{Trials} 2010; \textbf{11}. DOI:\href{https://doi.org/10.1186/1745-6215-11-85}{10.1186/1745-6215-11-85}.}

\leavevmode\vadjust pre{\hypertarget{ref-Thune2005}{}}%
\CSLLeftMargin{9 }%
\CSLRightInline{Thune JJ, Hoefsten DE, Lindholm MG, \emph{et al.} \href{https://doi.org/10.1161/circulationaha.105.558676}{Simple risk stratification at admission to identify patients with reduced mortality from primary angioplasty}. \emph{Circulation} 2005; \textbf{112}: 2017--21.}

\leavevmode\vadjust pre{\hypertarget{ref-Sussman2015}{}}%
\CSLLeftMargin{10 }%
\CSLRightInline{Sussman JB, Kent DM, Nelson JP, Hayward RA. \href{https://doi.org/10.1136/bmj.h454}{Improving diabetes prevention with benefit based tailored treatment: Risk based reanalysis of diabetes prevention program}. \emph{{BMJ}} 2015; \textbf{350}: h454--4.}

\leavevmode\vadjust pre{\hypertarget{ref-Kent2019}{}}%
\CSLLeftMargin{11 }%
\CSLRightInline{Kent DM, Paulus JK, Klaveren D van, \emph{et al.} \href{https://doi.org/10.7326/m18-3667}{The predictive approaches to treatment effect heterogeneity ({PATH}) statement}. \emph{Annals of Internal Medicine} 2019; \textbf{172}: 35.}

\leavevmode\vadjust pre{\hypertarget{ref-PathEnE}{}}%
\CSLLeftMargin{12 }%
\CSLRightInline{Kent DM, Klaveren D van, Paulus JK, \emph{et al.} \href{https://doi.org/10.7326/m18-3668}{The predictive approaches to treatment effect heterogeneity (PATH) statement: Explanation and elaboration}. \emph{Annals of Internal Medicine} 2020; \textbf{172}: W1--w25.}

\leavevmode\vadjust pre{\hypertarget{ref-hripcsak2015observational}{}}%
\CSLLeftMargin{13 }%
\CSLRightInline{Hripcsak G, Duke JD, Shah NH, \emph{et al.} Observational health data sciences and informatics (OHDSI): Opportunities for observational researchers. \emph{Studies in health technology and informatics} 2015; \textbf{216}: 574.}

\leavevmode\vadjust pre{\hypertarget{ref-Overhage2012}{}}%
\CSLLeftMargin{14 }%
\CSLRightInline{Overhage JM, Ryan PB, Reich CG, Hartzema AG, Stang PE. \href{https://doi.org/10.1136/amiajnl-2011-000376}{Validation of a common data model for active safety surveillance research}. \emph{Journal of the American Medical Informatics Association} 2012; \textbf{19}: 54--60.}

\leavevmode\vadjust pre{\hypertarget{ref-Ryan2013}{}}%
\CSLLeftMargin{15 }%
\CSLRightInline{Ryan PB, Schuemie MJ, Gruber S, Zorych I, Madigan D. \href{https://doi.org/10.1007/s40264-013-0099-6}{Empirical performance of a new user cohort method: Lessons for developing a risk identification and analysis system}. \emph{Drug Safety} 2013; \textbf{36}: 59--72.}

\leavevmode\vadjust pre{\hypertarget{ref-Suchard2019}{}}%
\CSLLeftMargin{16 }%
\CSLRightInline{Suchard MA, Schuemie MJ, Krumholz HM, \emph{et al.} \href{https://doi.org/10.1016/s0140-6736(19)32317-7}{Comprehensive comparative effectiveness and safety of first-line antihypertensive drug classes: A systematic, multinational, large-scale analysis}. \emph{The Lancet} 2019; \textbf{394}: 1816--26.}

\leavevmode\vadjust pre{\hypertarget{ref-Israili1992}{}}%
\CSLLeftMargin{17 }%
\CSLRightInline{Israili ZH. \href{https://doi.org/10.7326/0003-4819-117-3-234}{Cough and angioneurotic edema associated with angiotensin-converting enzyme inhibitor therapy}. \emph{Annals of Internal Medicine} 1992; \textbf{117}: 234.}

\leavevmode\vadjust pre{\hypertarget{ref-Rekkas2020}{}}%
\CSLLeftMargin{18 }%
\CSLRightInline{Rekkas A, Paulus JK, Raman G, \emph{et al.} Predictive approaches to heterogeneous treatment effects: A scoping review. \emph{{BMC} Medical Research Methodology} 2020; \textbf{20}. DOI:\href{https://doi.org/10.1186/s12874-020-01145-1}{10.1186/s12874-020-01145-1}.}

\leavevmode\vadjust pre{\hypertarget{ref-Varadhan2013}{}}%
\CSLLeftMargin{19 }%
\CSLRightInline{Varadhan R, Segal JB, Boyd CM, Wu AW, Weiss CO. \href{https://doi.org/10.1016/j.jclinepi.2013.02.009}{A framework for the analysis of heterogeneity of treatment effect in~patient-centered outcomes research}. \emph{Journal of Clinical Epidemiology} 2013; \textbf{66}: 818--25.}

\leavevmode\vadjust pre{\hypertarget{ref-Concato2000}{}}%
\CSLLeftMargin{20 }%
\CSLRightInline{Concato J, Shah N, Horwitz RI. \href{https://doi.org/10.1056/nejm200006223422507}{Randomized, controlled trials, observational studies, and the hierarchy of research designs}. \emph{New England Journal of Medicine} 2000; \textbf{342}: 1887--92.}

\leavevmode\vadjust pre{\hypertarget{ref-Ioannidis2001}{}}%
\CSLLeftMargin{21 }%
\CSLRightInline{Ioannidis JPA. \href{https://doi.org/10.1001/jama.286.7.821}{Comparison of evidence of treatment effects in randomized and nonrandomized studies}. \emph{{JAMA}} 2001; \textbf{286}: 821.}

\leavevmode\vadjust pre{\hypertarget{ref-Dahabreh2012}{}}%
\CSLLeftMargin{22 }%
\CSLRightInline{Dahabreh IJ, Sheldrick RC, Paulus JK, \emph{et al.} \href{https://doi.org/10.1093/eurheartj/ehs114}{Do observational studies using propensity score methods agree with randomized trials? A systematic comparison of studies on acute coronary syndromes}. \emph{European Heart Journal} 2012; \textbf{33}: 1893--901.}

\leavevmode\vadjust pre{\hypertarget{ref-Franklin2017}{}}%
\CSLLeftMargin{23 }%
\CSLRightInline{Franklin JM, Schneeweiss S. \href{https://doi.org/10.1002/cpt.857}{When and how can real world data analyses substitute for randomized controlled trials?} \emph{Clinical Pharmacology \& Therapeutics} 2017; \textbf{102}: 924--33.}

\leavevmode\vadjust pre{\hypertarget{ref-Anglemyer2014}{}}%
\CSLLeftMargin{24 }%
\CSLRightInline{Anglemyer A, Horvath HT, Bero L. Healthcare outcomes assessed with observational study designs compared with those assessed in randomized trials. \emph{Cochrane Database of Systematic Reviews} 2014; \textbf{2014}. DOI:\href{https://doi.org/10.1002/14651858.mr000034.pub2}{10.1002/14651858.mr000034.pub2}.}

\leavevmode\vadjust pre{\hypertarget{ref-Schuemie2018}{}}%
\CSLLeftMargin{25 }%
\CSLRightInline{Schuemie MJ, Ryan PB, Hripcsak G, Madigan D, Suchard MA. \href{https://doi.org/10.1098/rsta.2017.0356}{Improving reproducibility by using high-throughput observational studies with empirical calibration}. \emph{Philosophical Transactions of the Royal Society A: Mathematical, Physical and Engineering Sciences} 2018; \textbf{376}: 20170356.}

\leavevmode\vadjust pre{\hypertarget{ref-Dahabreh2016}{}}%
\CSLLeftMargin{26 }%
\CSLRightInline{Dahabreh IJ, Hayward R, Kent DM. \href{https://doi.org/10.1093/ije/dyw125}{Using group data to treat individuals: Understanding heterogeneous treatment effects in the age of precision medicine and patient-centred evidence}. \emph{International Journal of Epidemiology} 2016; \textbf{45}: 2184--93.}

\leavevmode\vadjust pre{\hypertarget{ref-Reps2018}{}}%
\CSLLeftMargin{27 }%
\CSLRightInline{Reps JM, Schuemie MJ, Suchard MA, Ryan PB, Rijnbeek PR. \href{https://doi.org/10.1093/jamia/ocy032}{Design and implementation of a standardized framework to generate and evaluate patient-level prediction models using observational healthcare data}. \emph{Journal of the American Medical Informatics Association} 2018; \textbf{25}: 969--75.}

\leavevmode\vadjust pre{\hypertarget{ref-Collins2015}{}}%
\CSLLeftMargin{28 }%
\CSLRightInline{Collins GS, Reitsma JB, Altman DG, Moons K. \href{https://doi.org/10.1186/s12916-014-0241-z}{Transparent reporting of a multivariable prediction model for individual prognosis or diagnosis ({TRIPOD}): The {TRIPOD} statement}. \emph{{BMC} Medicine} 2015; \textbf{13}: 1.}

\leavevmode\vadjust pre{\hypertarget{ref-Moons2015}{}}%
\CSLLeftMargin{29 }%
\CSLRightInline{Moons KGM, Altman DG, Reitsma JB, \emph{et al.} \href{https://doi.org/10.7326/m14-0698}{Transparent reporting of a multivariable prediction model for individual prognosis or diagnosis ({TRIPOD}): Explanation and elaboration}. \emph{Annals of Internal Medicine} 2015; \textbf{162}: W1.}

\leavevmode\vadjust pre{\hypertarget{ref-Burke2014}{}}%
\CSLLeftMargin{30 }%
\CSLRightInline{Burke JF, Hayward RA, Nelson JP, Kent DM. \href{https://doi.org/10.1161/circoutcomes.113.000497}{Using internally developed risk models to assess heterogeneity in treatment effects in clinical trials}. \emph{Circulation: Cardiovascular Quality and Outcomes} 2014; \textbf{7}: 163--9.}

\leavevmode\vadjust pre{\hypertarget{ref-vanKlaveren2019}{}}%
\CSLLeftMargin{31 }%
\CSLRightInline{Klaveren D van, Balan TA, Steyerberg EW, Kent DM. \href{https://doi.org/10.1016/j.jclinepi.2019.05.029}{Models with interactions overestimated heterogeneity of treatment effects and were prone to treatment mistargeting}. \emph{Journal of Clinical Epidemiology} 2019; \textbf{114}: 72--83.}

\leavevmode\vadjust pre{\hypertarget{ref-Austin2011}{}}%
\CSLLeftMargin{32 }%
\CSLRightInline{Austin PC. \href{https://doi.org/10.1080/00273171.2011.568786}{An introduction to propensity score methods for reducing the effects of confounding in observational studies}. \emph{Multivariate Behavioral Research} 2011; \textbf{46}: 399--424.}

\leavevmode\vadjust pre{\hypertarget{ref-Schuemie2014}{}}%
\CSLLeftMargin{33 }%
\CSLRightInline{Schuemie MJ, Ryan PB, DuMouchel W, Suchard MA, Madigan D. \href{https://doi.org/10.1002/sim.5925}{Interpreting observational studies: Why empirical calibration is needed to correct p-values}. \emph{Statistics in Medicine} 2014; \textbf{33}: 209--18.}

\leavevmode\vadjust pre{\hypertarget{ref-Schuemie2571}{}}%
\CSLLeftMargin{34 }%
\CSLRightInline{Schuemie MJ, Hripcsak G, Ryan PB, Madigan D, Suchard MA. \href{https://doi.org/10.1073/pnas.1708282114}{Empirical confidence interval calibration for population-level effect estimation studies in observational healthcare data}. \emph{Proceedings of the National Academy of Sciences} 2018; \textbf{115}: 2571--7.}

\end{CSLReferences}

\setlength{\parindent}{0in}
\setlength{\leftskip}{0in}

\noindent

\newpage

\hypertarget{acknowledgements}{%
\section{ACKNOWLEDGEMENTS}\label{acknowledgements}}

AR and PRR have received funding from the Innovative Medicines Initiative 2
Joint Undertaking (JU) under grant agreement No 80696. The JU receives support
from the European Union's Horizon 2020 research and innovation programme and
EFPIA.

\hypertarget{author-contributions}{%
\section{AUTHOR CONTRIBUTIONS}\label{author-contributions}}

AR, PRR, and DVK conceptualised the study. AR, DVK, EWS, and DMK developed the
methodology. PBR, and PRR acquired the data and AR analysed the data. AR
developed the software and wrote drafted the manuscript, which was critically
reviewed by DVK, PBR, EWS, DMK, and PRR. AR, PBR, and PRR had full access to the
raw data. All authors read and approved the manuscript and had final
responsibility for the decision to submit for publication.

\hypertarget{competing-interests}{%
\section{COMPETING INTERESTS}\label{competing-interests}}

AR and PRR work for a group that received unconditional research grants from
Boehringer-Ingelheim, GSK, Janssen Research \& Development, Novartis, Pfizer,
Yamanouchi, and Servier. None of these grants result in a conflict of interest
for the content of this paper. PBR is an employee of Janssen R\&D, subsidiary of
Johnson \& Johnson. DVK, DMK, EWS have nothing to declare.

\ifarXiv
    

\section*{Supplementary material}

**Supplementary material**



# CONTENTS



**COHORT DEFINITIONS**

**Treatment cohorts**

 First-line new user ACE inhibitors

*Initial Event Cohort*

People having any of the following:

- a drug exposure of ACE inhibitors (see 2 below)
    - for the first time in the person's history

with continuous observation of at least 365 days prior and 0 days after event index date, and limit initial events to: earliest event per person.

For people matching the Primary Events, include:

Having all of the following criteria:

- exactly 0 occurrences of a drug exposure of Hypertension drugs (see 4 below) where event starts between all days Before and 1 days Before index start date
- and at least 1 occurrences of a condition occurrence of Hypertensive disorder (see 1 below) where event starts between 365 days Before and 0 days After index start date
- and exactly 1 distinct occurrences of a drug era of [Hypertension drugs (see 4 below) where event starts between 0 days Before and 7 days After index start date

Limit cohort of initial events to: earliest event per person.

Limit qualifying cohort to: earliest event per person.

*End Date Strategy*

Custom Drug Era Exit Criteria
This strategy creates a drug era from the codes found in the specified concept set. If the index event is found within an era, the cohort end date will use the era's end date. Otherwise, it will use the observation period end date that contains the index event.

Use the era end date of ACE inhibitors (see 2 below)

- allowing 30 days between exposures
- adding 0 days after exposure end
- using days supply and exposure end date for exposure duration.

Cohort Collapse Strategy:
Collapse cohort by era with a gap size of 0 days.



*Concept Set Definitions*

1. Hypertensive disorder

| Concept Id | Concept Name | Domain | Vocabulary | Excluded | Descendants | Mapped |
|---|---|---|---|---|---|---|
| 316866 | Hypertensive disorder | Condition | SNOMED | NO | YES | NO |

2. ACE inhibitors

| Concept Id | Concept Name | Domain | Vocabulary | Excluded | Descendants | Mapped |
|---|---|---|---|---|---|---|
| 1308216 | Lisinopril | Drug | RxNorm | NO | YES | NO |
| 1310756 | moexipril | Drug | RxNorm | NO | YES | NO |
| 1331235 | quinapril | Drug | RxNorm | NO | YES | NO |
| 1334456 | Ramipril | Drug | RxNorm | NO | YES | NO |
| 1335471 | benazepril | Drug | RxNorm | NO | YES | NO |
| 1340128 | Captopril | Drug | RxNorm | NO | YES | NO |
| 1341927 | Enalapril | Drug | RxNorm | NO | YES | NO |
| 1342439 | trandolapril | Drug | RxNorm | NO | YES | NO |
| 1363749 | Fosinopril | Drug | RxNorm | NO | YES | NO |
| 1373225 | Perindopril | Drug | RxNorm | NO | YES | NO |



3. First-line hypertension drugs

| Concept Id | Concept Name | Domain | Vocabulary | Excluded | Descendants | Mapped |
|---|---|---|---|---|---|---|
| 907013 | Metolazone | Drug | RxNorm | NO | YES | NO |
| 974166 | Hydrochlorothiazide | Drug | RxNorm | NO | YES | NO |
| 978555 | Indapamide | Drug | RxNorm | NO | YES | NO |
| 1307863 | Verapamil | Drug | RxNorm | NO | YES | NO |
| 1308216 | Lisinopril | Drug | RxNorm | NO | YES | NO |
| 1308842 | valsartan | Drug | RxNorm | NO | YES | NO |
| 1310756 | moexipril | Drug | RxNorm | NO | YES | NO |
| 1317640 | telmisartan | Drug | RxNorm | NO | YES | NO |
| 1318137 | Nicardipine | Drug | RxNorm | NO | YES | NO |
| 1318853 | Nifedipine | Drug | RxNorm | NO | YES | NO |
| 1319880 | Nisoldipine | Drug | RxNorm | NO | YES | NO |
| 1326012 | Isradipine | Drug | RxNorm | NO | YES | NO |
| 1328165 | Diltiazem | Drug | RxNorm | NO | YES | NO |
| 1331235 | quinapril | Drug | RxNorm | NO | YES | NO |
| 1332418 | Amlodipine | Drug | RxNorm | NO | YES | NO |
| 1334456 | Ramipril | Drug | RxNorm | NO | YES | NO |
| 1335471 | benazepril | Drug | RxNorm | NO | YES | NO |
| 1340128 | Captopril | Drug | RxNorm | NO | YES | NO |
| 1341927 | Enalapril | Drug | RxNorm | NO | YES | NO |
| 1342439 | trandolapril | Drug | RxNorm | NO | YES | NO |



| 1346686 | eprosartan | Drug | RxNorm | NO | YES | NO |
| 1347384 | irbesartan | Drug | RxNorm | NO | YES | NO |
| 1351557 | candesartan | Drug | RxNorm | NO | YES | NO |
| 1353776 | Felodipine | Drug | RxNorm | NO | YES | NO |
| 1363749 | Fosinopril | Drug | RxNorm | NO | YES | NO |
| 1367500 | Losartan | Drug | RxNorm | NO | YES | NO |
| 1373225 | Perindopril | Drug | RxNorm | NO | YES | NO |
| 1395058 | Chlorthalidone | Drug | RxNorm | NO | YES | NO |
| 40226742 | olmesartan | Drug | RxNorm | NO | YES | NO |
| 40235485 | azilsartan | Drug | RxNorm | NO | YES | NO |

4. Hypertension drugs

| Concept Id | Concept Name | Domain | Vocabulary | Excluded | Descendants | Mapped |
| --- | --- | --- | --- | --- | --- | --- |
| 40235485 | azilsartan | Drug | RxNorm | NO | YES | NO |
| 40226742 | olmesartan | Drug | RxNorm | NO | YES | NO |
| 1398937 | Clonidine | Drug | RxNorm | NO | YES | NO |
| 1395058 | Chlorthalidone | Drug | RxNorm | NO | YES | NO |
| 1386957 | Labetalol | Drug | RxNorm | NO | YES | NO |
| 1373928 | Hydralazine | Drug | RxNorm | NO | YES | NO |
| 1373225 | Perindopril | Drug | RxNorm | NO | YES | NO |
| 1367500 | Losartan | Drug | RxNorm | NO | YES | NO |
| 1363749 | Fosinopril | Drug | RxNorm | NO | YES | NO |
| 1363053 | Doxazosin | Drug | RxNorm | NO | YES | NO |



| 1353776 | Felodipine | Drug | RxNorm | NO | YES | NO |
|---|---|---|---|---|---|---|
| 1353766 | Propranolol | Drug | RxNorm | NO | YES | NO |
| 1351557 | candesartan | Drug | RxNorm | NO | YES | NO |
| 1350489 | Prazosin | Drug | RxNorm | NO | YES | NO |
| 1347384 | irbesartan | Drug | RxNorm | NO | YES | NO |
| 1346823 | carvedilol | Drug | RxNorm | NO | YES | NO |
| 1346686 | eprosartan | Drug | RxNorm | NO | YES | NO |
| 1345858 | Pindolol | Drug | RxNorm | NO | YES | NO |
| 1344965 | Guanfacine | Drug | RxNorm | NO | YES | NO |
| 1342439 | trandolapril | Drug | RxNorm | NO | YES | NO |
| 1341927 | Enalapril | Drug | RxNorm | NO | YES | NO |
| 1341238 | Terazosin | Drug | RxNorm | NO | YES | NO |
| 1340128 | Captopril | Drug | RxNorm | NO | YES | NO |
| 1338005 | Bisoprolol | Drug | RxNorm | NO | YES | NO |
| 1335471 | benazepril | Drug | RxNorm | NO | YES | NO |
| 1334456 | Ramipril | Drug | RxNorm | NO | YES | NO |
| 1332418 | Amlodipine | Drug | RxNorm | NO | YES | NO |
| 1331235 | quinapril | Drug | RxNorm | NO | YES | NO |
| 1328165 | Diltiazem | Drug | RxNorm | NO | YES | NO |
| 1327978 | Penbutolol | Drug | RxNorm | NO | YES | NO |
| 1326012 | Isradipine | Drug | RxNorm | NO | YES | NO |
| 1322081 | Betaxolol | Drug | RxNorm | NO | YES | NO |
| 1319998 | Acebutolol | Drug | RxNorm | NO | YES | NO |



| 1319880 | Nisoldipine | Drug | RxNorm | NO | YES | NO |
|---------|-------------|------|--------|-----|-----|-----|
| 1318853 | Nifedipine | Drug | RxNorm | NO | YES | NO |
| 1318137 | Nicardipine | Drug | RxNorm | NO | YES | NO |
| 1317967 | aliskiren | Drug | RxNorm | NO | YES | NO |
| 1317640 | telmisartan | Drug | RxNorm | NO | YES | NO |
| 1314577 | nebivolol | Drug | RxNorm | NO | YES | NO |
| 1314002 | Atenolol | Drug | RxNorm | NO | YES | NO |
| 1313200 | Nadolol | Drug | RxNorm | NO | YES | NO |
| 1310756 | moexipril | Drug | RxNorm | NO | YES | NO |
| 1309799 | eplerenone | Drug | RxNorm | NO | YES | NO |
| 1309068 | Minoxidil | Drug | RxNorm | NO | YES | NO |
| 1308842 | valsartan | Drug | RxNorm | NO | YES | NO |
| 1308216 | Lisinopril | Drug | RxNorm | NO | YES | NO |
| 1307863 | Verapamil | Drug | RxNorm | NO | YES | NO |
| 1307046 | Metoprolol | Drug | RxNorm | NO | YES | NO |
| 1305447 | Methyldopa | Drug | RxNorm | NO | YES | NO |
| 991382 | Amiloride | Drug | RxNorm | NO | YES | NO |
| 978555 | Indapamide | Drug | RxNorm | NO | YES | NO |
| 974166 | Hydrochlorothiazide | Drug | RxNorm | NO | YES | NO |
| 970250 | Spironolactone | Drug | RxNorm | NO | YES | NO |
| 956874 | Furosemide | Drug | RxNorm | NO | YES | NO |
| 942350 | torsemide | Drug | RxNorm | NO | YES | NO |
| 932745 | Bumetanide | Drug | RxNorm | NO | YES | NO |



| 907013 | Metolazone | Drug | RxNorm | NO | YES | NO |
|--------|------------|------|--------|-----|-----|-----|
| 904542 | Triamterene | Drug | RxNorm | NO | YES | NO |

First-line new user beta blockers

*Initial Event Cohort*

People having any of the following:

- a drug exposure of Beta blockers (see 2 below)
  - ♦ for the first time in the person's history

with continuous observation of at least 365 days prior and 0 days after event index date, and limit initial events to: earliest event per person.

For people matching the Primary Events, include:

Having all of the following criteria:

- exactly 0 occurrences of a drug exposure of Hypertension drugs (see 4 below) where event starts between all days Before and 1 days Before index start date
- and at least 1 occurrences of a condition occurrence of Hypertensive disorder (see 1 below) where event starts between 365 days Before and 0 days After index start date
- and exactly 1 distinct occurrences of a drug era of Hypertension drugs (see 4 below) where event starts between 0 days Before and 7 days After index start date

Limit cohort of initial events to: earliest event per person.

Limit qualifying cohort to: earliest event per person.

*End Date Strategy*

Custom Drug Era Exit Criteria
This strategy creates a drug era from the codes found in the specified concept set. If the index event is found within an era, the cohort end date will use the era's end date. Otherwise, it will use the observation period end date that contains the index event.

Use the era end date of Beta blockers (see 2 below)

- allowing 30 days between exposures
- adding 0 days after exposure end

Cohort Collapse Strategy:
Collapse cohort by era with a gap size of 0 days.



*Concept Set Definitions*

1. Hypertensive disorder

| Concept Id | Concept Name | Domain | Vocabulary | Excluded | Descendants | Mapped |
|---|---|---|---|---|---|---|
| 316866 | Hypertensive disorder | Condition | SNOMED | NO | YES | NO |

2. Beta blockers

| Concept Id | Concept Name | Domain | Vocabulary | Excluded | Descendants | Mapped |
|---|---|---|---|---|---|---|
| 1307046 | Metoprolol | Drug | RxNorm | NO | YES | NO |
| 1313200 | Nadolol | Drug | RxNorm | NO | YES | NO |
| 1314002 | Atenolol | Drug | RxNorm | NO | YES | NO |
| 1314577 | nebivolol | Drug | RxNorm | NO | YES | NO |
| 1319998 | Acebutolol | Drug | RxNorm | NO | YES | NO |
| 1322081 | Betaxolol | Drug | RxNorm | NO | YES | NO |
| 1327978 | Penbutolol | Drug | RxNorm | NO | YES | NO |
| 1338005 | Bisoprolol | Drug | RxNorm | NO | YES | NO |
| 1345858 | Pindolol | Drug | RxNorm | NO | YES | NO |
| 1346823 | carvedilol | Drug | RxNorm | NO | YES | NO |
| 1353766 | Propranolol | Drug | RxNorm | NO | YES | NO |
| 1386957 | Labetalol | Drug | RxNorm | NO | YES | NO |

3. First-line hypertension drugs



| Concept Id | Concept Name | Domain | Vocabulary | Excluded | Descendants | Mapped |
|---|---|---|---|---|---|---|
| 907013 | Metolazone | Drug | RxNorm | NO | YES | NO |
| 974166 | Hydrochlorothiazide | Drug | RxNorm | NO | YES | NO |
| 978555 | Indapamide | Drug | RxNorm | NO | YES | NO |
| 1307863 | Verapamil | Drug | RxNorm | NO | YES | NO |
| 1308216 | Lisinopril | Drug | RxNorm | NO | YES | NO |
| 1308842 | valsartan | Drug | RxNorm | NO | YES | NO |
| 1310756 | moexipril | Drug | RxNorm | NO | YES | NO |
| 1317640 | telmisartan | Drug | RxNorm | NO | YES | NO |
| 1318137 | Nicardipine | Drug | RxNorm | NO | YES | NO |
| 1318853 | Nifedipine | Drug | RxNorm | NO | YES | NO |
| 1319880 | Nisoldipine | Drug | RxNorm | NO | YES | NO |
| 1326012 | Isradipine | Drug | RxNorm | NO | YES | NO |
| 1328165 | Diltiazem | Drug | RxNorm | NO | YES | NO |
| 1331235 | quinapril | Drug | RxNorm | NO | YES | NO |
| 1332418 | Amlodipine | Drug | RxNorm | NO | YES | NO |
| 1334456 | Ramipril | Drug | RxNorm | NO | YES | NO |
| 1335471 | benazepril | Drug | RxNorm | NO | YES | NO |
| 1340128 | Captopril | Drug | RxNorm | NO | YES | NO |
| 1341927 | Enalapril | Drug | RxNorm | NO | YES | NO |
| 1342439 | trandolapril | Drug | RxNorm | NO | YES | NO |
| 1346686 | eprosartan | Drug | RxNorm | NO | YES | NO |



| 1347384 | irbesartan | Drug | RxNorm | NO | YES | NO |
|---|---|---|---|---|---|---|
| 1351557 | candesartan | Drug | RxNorm | NO | YES | NO |
| 1353776 | Felodipine | Drug | RxNorm | NO | YES | NO |
| 1363749 | Fosinopril | Drug | RxNorm | NO | YES | NO |
| 1367500 | Losartan | Drug | RxNorm | NO | YES | NO |
| 1373225 | Perindopril | Drug | RxNorm | NO | YES | NO |
| 1395058 | Chlorthalidone | Drug | RxNorm | NO | YES | NO |
| 40226742 | olmesartan | Drug | RxNorm | NO | YES | NO |
| 40235485 | azilsartan | Drug | RxNorm | NO | YES | NO |

4. Hypertension drugs

| Concept Id | Concept Name | Domain | Vocabulary | Excluded | Descendants | Mapped |
|---|---|---|---|---|---|---|
| 904542 | Triamterene | Drug | RxNorm | NO | YES | NO |
| 907013 | Metolazone | Drug | RxNorm | NO | YES | NO |
| 932745 | Bumetanide | Drug | RxNorm | NO | YES | NO |
| 942350 | torsemide | Drug | RxNorm | NO | YES | NO |
| 956874 | Furosemide | Drug | RxNorm | NO | YES | NO |
| 970250 | Spironolactone | Drug | RxNorm | NO | YES | NO |
| 974166 | Hydrochlorothiazide | Drug | RxNorm | NO | YES | NO |
| 978555 | Indapamide | Drug | RxNorm | NO | YES | NO |
| 991382 | Amiloride | Drug | RxNorm | NO | YES | NO |
| 1305447 | Methyldopa | Drug | RxNorm | NO | YES | NO |
| 1307046 | Metoprolol | Drug | RxNorm | NO | YES | NO |



| 1307863 | Verapamil | Drug | RxNorm | NO | YES | NO |
|---------|-----------|------|--------|----|----|----|
| 1308216 | Lisinopril | Drug | RxNorm | NO | YES | NO |
| 1308842 | valsartan | Drug | RxNorm | NO | YES | NO |
| 1309068 | Minoxidil | Drug | RxNorm | NO | YES | NO |
| 1309799 | eplerenone | Drug | RxNorm | NO | YES | NO |
| 1310756 | moexipril | Drug | RxNorm | NO | YES | NO |
| 1313200 | Nadolol | Drug | RxNorm | NO | YES | NO |
| 1314002 | Atenolol | Drug | RxNorm | NO | YES | NO |
| 1314577 | nebivolol | Drug | RxNorm | NO | YES | NO |
| 1317640 | telmisartan | Drug | RxNorm | NO | YES | NO |
| 1317967 | aliskiren | Drug | RxNorm | NO | YES | NO |
| 1318137 | Nicardipine | Drug | RxNorm | NO | YES | NO |
| 1318853 | Nifedipine | Drug | RxNorm | NO | YES | NO |
| 1319880 | Nisoldipine | Drug | RxNorm | NO | YES | NO |
| 1319998 | Acebutolol | Drug | RxNorm | NO | YES | NO |
| 1322081 | Betaxolol | Drug | RxNorm | NO | YES | NO |
| 1326012 | Isradipine | Drug | RxNorm | NO | YES | NO |
| 1327978 | Penbutolol | Drug | RxNorm | NO | YES | NO |
| 1328165 | Diltiazem | Drug | RxNorm | NO | YES | NO |
| 1331235 | quinapril | Drug | RxNorm | NO | YES | NO |
| 1332418 | Amlodipine | Drug | RxNorm | NO | YES | NO |
| 1334456 | Ramipril | Drug | RxNorm | NO | YES | NO |
| 1335471 | benazepril | Drug | RxNorm | NO | YES | NO |



| | | | | | | |
|---|---|---|---|---|---|---|
| 1338005 | Bisoprolol | Drug | RxNorm | NO | YES | NO |
| 1340128 | Captopril | Drug | RxNorm | NO | YES | NO |
| 1341238 | Terazosin | Drug | RxNorm | NO | YES | NO |
| 1341927 | Enalapril | Drug | RxNorm | NO | YES | NO |
| 1342439 | trandolapril | Drug | RxNorm | NO | YES | NO |
| 1344965 | Guanfacine | Drug | RxNorm | NO | YES | NO |
| 1345858 | Pindolol | Drug | RxNorm | NO | YES | NO |
| 1346686 | eprosartan | Drug | RxNorm | NO | YES | NO |
| 1346823 | carvedilol | Drug | RxNorm | NO | YES | NO |
| 1347384 | irbesartan | Drug | RxNorm | NO | YES | NO |
| 1350489 | Prazosin | Drug | RxNorm | NO | YES | NO |
| 1351557 | candesartan | Drug | RxNorm | NO | YES | NO |
| 1353766 | Propranolol | Drug | RxNorm | NO | YES | NO |
| 1353776 | Felodipine | Drug | RxNorm | NO | YES | NO |
| 1363053 | Doxazosin | Drug | RxNorm | NO | YES | NO |
| 1363749 | Fosinopril | Drug | RxNorm | NO | YES | NO |
| 1367500 | Losartan | Drug | RxNorm | NO | YES | NO |
| 1373225 | Perindopril | Drug | RxNorm | NO | YES | NO |
| 1373928 | Hydralazine | Drug | RxNorm | NO | YES | NO |
| 1386957 | Labetalol | Drug | RxNorm | NO | YES | NO |
| 1395058 | Chlorthalidone | Drug | RxNorm | NO | YES | NO |
| 1398937 | Clonidine | Drug | RxNorm | NO | YES | NO |
| 40226742 | olmesartan | Drug | RxNorm | NO | YES | NO |



| 40235485 | azilsartan | Drug | RxNorm | NO | YES | NO |
| --- | --- | --- | --- | --- | --- | --- |

**Outcome cohorts**

Hospitalization with heart failure

Inpatient or ER visits with heart failure condition record; all qualifying inpatient visits occurring > 7 days apart are considered independent episodes

*Initial Event Cohort*

People having any of the following:

- a visit occurrence of Inpatient or ER visit (see 1 below)
    having one of the following:

    ♦ at least 1 occurrences of a condition occurrence of Heart Failure (see 2 below) where event starts between 0 days Before and all days After index start date and event starts between all days Before and 0 days After index end date

with continuous observation of at least 0 days prior and 0 days after event index date, and limit initial events to: all events per person.

Limit qualifying cohort to: all events per person.

*End Date Strategy*

Date Offset Exit Criteria
This cohort definition end date will be the index event's end date plus 0 days

Cohort Collapse Strategy:
Collapse cohort by era with a gap size of 7 days.

*Concept Set Definitions*

1. Inpatient or ER visit

| Concept Id | Concept Name | Domain | Vocabulary | Excluded | Descendants | Mapped |
| --- | --- | --- | --- | --- | --- | --- |
| 262 | Emergency Room and Inpatient Visit | Visit | Visit | NO | YES | NO |
| 9201 | Inpatient Visit | Visit | Visit | NO | YES | NO |
| 9203 | Emergency Room Visit | Visit | Visit | NO | YES | NO |



2. Heart Failure

| Concept Id | Concept Name | Domain | Vocabulary | Excluded | Descendants | Mapped |
|---|---|---|---|---|---|---|
| 315295 | Congestive rheumatic heart failure | Condition | SNOMED | YES | YES | NO |
| 316139 | Heart failure | Condition | SNOMED | NO | YES | NO |

Acute Myocardial Infarction

*Initial Event Cohort*

People having any of the following:

- a condition occurrence of Acute myocardial Infarction (see 2 below)

with continuous observation of at least 0 days prior and 0 days after event index date, and limit initial events to: all events per person.

For people matching the Primary Events, include:

Having any of the following criteria:

- at least 1 occurrences of a visit occurrence of Inpatient or ER visit (see 1 below) where event starts between all days Before and 0 days After index start date and event ends between 0 days Before and all days After index start date

Limit cohort of initial events to: all events per person.

Limit qualifying cohort to: all events per person.

*End Date Strategy*

Date Offset Exit Criteria
This cohort definition end date will be the index event's start date plus 7 days

Cohort Collapse Strategy:
Collapse cohort by era with a gap size of 180 days.

*Concept Set Definitions*

1. Inpatient or ER visit



| Concept Id | Concept Name | Domain | Vocabulary | Excluded | Descendants | Mapped |
|---|---|---|---|---|---|---|
| 262 | Emergency Room and Inpatient Visit | Visit | Visit | NO | YES | NO |
| 9201 | Inpatient Visit | Visit | Visit | NO | YES | NO |
| 9203 | Emergency Room Visit | Visit | Visit | NO | YES | NO |

2. Acute myocardial Infarction

| Concept Id | Concept Name | Domain | Vocabulary | Excluded | Descendants | Mapped |
|---|---|---|---|---|---|---|
| 314666 | Old myocardial infarction | Condition | SNOMED | YES | YES | NO |
| 4329847 | Myocardial infarction | Condition | SNOMED | NO | YES | NO |

Stroke (ischemic or hemorrhagic) events

Stroke (ischemic or hemorrhagic) condition record during an inpatient or ER visit; successive records with > 180 day gap are considered independent episodes

*Initial Event Cohort*

People having any of the following:

• a condition occurrence of Stroke ischemic or hemorrhagic (see 2 below)

with continuous observation of at least 0 days prior and 0 days after event index date, and limit initial events to: all events per person.

For people matching the Primary Events, include:

Having any of the following criteria:

• at least 1 occurrences of a visit occurrence of Inpatient or ER visit (see 1 below)

where event starts between all days Before and 1 days After index start date and event ends between 0 days Before and all days After index start date

Limit cohort of initial events to: all events per person.



Limit qualifying cohort to: all events per person.

*End Date Strategy*

Date Offset Exit Criteria
This cohort definition end date will be the index event's start date plus 7 days

Cohort Collapse Strategy:
Collapse cohort by era with a gap size of 180 days.

*Concept Set Definitions*

1. Inpatient or ER visit

| Concept Id | Concept Name | Domain | Vocabulary | Excluded | Descendants | Mapped |
|---|---|---|---|---|---|---|
| 262 | Emergency Room and Inpatient Visit | Visit | Visit | NO | YES | NO |
| 9201 | Inpatient Visit | Visit | Visit | NO | YES | NO |
| 9203 | Emergency Room Visit | Visit | Visit | NO | YES | NO |

2. Stroke (ischemic or hemorrhagic)

| Concept Id | Concept Name | Domain | Vocabulary | Excluded | Descendants | Mapped |
|---|---|---|---|---|---|---|
| 372924 | Cerebral artery occlusion | Condition | SNOMED | NO | NO | NO |
| 375557 | Cerebral embolism | Condition | SNOMED | NO | NO | NO |
| 376713 | Cerebral haemorrhage | Condition | SNOMED | NO | NO | NO |
| 432923 | Subarachnoid haemorrhage | Condition | SNOMED | NO | NO | NO |



| 439847 | Intracranial haemorrhage | Condition | SNOMED | NO | NO | NO |
|---|---|---|---|---|---|---|
| 441874 | Cerebral thrombosis | Condition | SNOMED | NO | NO | NO |
| 443454 | Cerebral infarction | Condition | SNOMED | NO | YES | NO |

Abnormal weight gain events

Abnormal weight gain record of any type; successive records with > 90 day gap are considered independent episodes

*Initial Event Cohort*

People having any of the following:

- an observation of Abnormal weight gain (see 1 below)

with continuous observation of at least 0 days prior and 0 days after event index date, and limit initial events to: all events per person.

Limit qualifying cohort to: all events per person.

*End Date Strategy*

Date Offset Exit Criteria
This cohort definition end date will be the index event's start date plus 1 days

Cohort Collapse Strategy:
Collapse cohort by era with a gap size of 90 days.

*Concept Set Definitions*

1. Abnormal weight gain

| Concept Id | Concept Name | Domain | Vocabulary | Excluded | Descendants | Mapped |
|---|---|---|---|---|---|---|
| 439141 | Abnormal weight gain | Observation | SNOMED | NO | YES | NO |

Angioedema events



Angioedema condition record during an inpatient or ER visit; successive records with >7 day gap are considered independent episodes

*Initial Event Cohort*

People having any of the following:

• a condition occurrence of Angioedema (see 2 below)

with continuous observation of at least 0 days prior and 0 days after event index date, and limit initial events to: all events per person.

For people matching the Primary Events, include:

Having any of the following criteria:

• at least 1 occurrences of a visit occurrence of Inpatient or ER visit (see 1 below)

where event starts between all days Before and 0 days After index start date and event ends between 0 days Before and all days After index start date

Limit cohort of initial events to: all events per person.

Limit qualifying cohort to: all events per person.

*End Date Strategy*

Date Offset Exit Criteria
This cohort definition end date will be the index event's start date plus 7 days

Cohort Collapse Strategy:
Collapse cohort by era with a gap size of 30 days.



*Concept Set Definitions*

1. Inpatient or ER visit

| Concept Id | Concept Name | Domain | Vocabulary | Excluded | Descendants | Mapped |
|---|---|---|---|---|---|---|
| 262 | Emergency Room and Inpatient Visit | Visit | Visit | NO | YES | NO |
| 9201 | Inpatient Visit | Visit | Visit | NO | YES | NO |
| 9203 | Emergency Room Visit | Visit | Visit | NO | YES | NO |

2. Angioedema

| Concept Id | Concept Name | Domain | Vocabulary | Excluded | Descendants | Mapped |
|---|---|---|---|---|---|---|
| 432791 | Angioedema | Condition | SNOMED | NO | YES | NO |

Cough events

Cough condition record of any type; successive records with > 90 day gap are considered independent episodes

*Initial Event Cohort*

People having any of the following:

• a condition occurrence of Cough (see 1 below)

with continuous observation of at least 0 days prior and 0 days after event index date, and limit initial events to: all events per person.

Limit qualifying cohort to: all events per person.

*End Date Strategy*

Date Offset Exit Criteria
This cohort definition end date will be the index event's start date plus 1 days

Cohort Collapse Strategy:
Collapse cohort by era with a gap size of 90 days.



Concept Set Definitions

1. Cough

| Concept Id | Concept Name | Domain | Vocabulary | Excluded | Descendants | Mapped |
|---|---|---|---|---|---|---|
| 254761 | Cough | Condition | SNOMED | NO | YES | NO |

Hyperkalemia events

Condition record for hyperkalemia or potassium measurements > 5.6 mmol/L; successive records with >90 day gap are considered independent episodes

*Initial Event Cohort*

People having any of the following:

- a condition occurrence of Hyperkalemia (see 1 below)
- a measurement of Potassium measurement (see 2 below)
    - with value as number > 5.6
    - unit is any of: millimole per litre

with continuous observation of at least 0 days prior and 0 days after event index date, and limit initial events to: all events per person.

Limit qualifying cohort to: all events per person.

*End Date Strategy*

Date Offset Exit Criteria
This cohort definition end date will be the index event's start date plus 1 days

Cohort Collapse Strategy:
Collapse cohort by era with a gap size of 90 days.

*Concept Set Definitions*

1. Hyperkalemia

| Concept Id | Concept Name | Domain | Vocabulary | Excluded | Descendants | Mapped |
|---|---|---|---|---|---|---|
| 434610 | Hyperkalemia | Condition | SNOMED | NO | YES | NO |



2. Potassium measurement

| Concept Id | Concept Name | Domain | Vocabulary | Excluded | Descendants | Mapped |
|---|---|---|---|---|---|---|
| 4245152 | Potassium measurement | Measurement | SNOMED | NO | YES | NO |
| 4276440 | Potassium level - finding | Condition | SNOMED | NO | YES | NO |
| 40789893 | Potassium | Bld-Ser-Plas | Measurement | LOINC | NO | YES | NO |

Hypokalemia events

Hypokalemia condition record of any type; successive records with > 90 day gap are considered independent episodes

*Initial Event Cohort*

People having any of the following:

•   a condition occurrence of Hypokalemia (see 1 below)

with continuous observation of at least 0 days prior and 0 days after event index date, and limit initial events to: all events per person.

Limit qualifying cohort to: all events per person.

*End Date Strategy*

Date Offset Exit Criteria
This cohort definition end date will be the index event's start date plus 1 days

Cohort Collapse Strategy:
Collapse cohort by era with a gap size of 90 days.

*Concept Set Definitions*

1. Hypokalemia

| Concept Id | Concept Name | Domain | Vocabulary | Excluded | Descendants | Mapped |
|---|---|---|---|---|---|---|
| 437833 | Hypokalemia | Condition | SNOMED | NO | YES | NO |
| 45769152 | Bartter syndrome | Condition | SNOMED | YES | YES | NO |



Hypotension events

Hypotension condition record of any type; successive records with > 90 day gap are considered independent episodes

*Initial Event Cohort*

People having any of the following:

• a condition occurrence of Hypotension (see 1 below)

with continuous observation of at least 0 days prior and 0 days after event index date, and limit initial events to: all events per person.

Limit qualifying cohort to: all events per person.

*End Date Strategy*

Date Offset Exit Criteria
This cohort definition end date will be the index event's start date plus 1 days

Cohort Collapse Strategy:
Collapse cohort by era with a gap size of 90 days.

*Concept Set Definitions*

1. Hypotension

| Concept Id | Concept Name | Domain | Vocabulary | Excluded | Descendants | Mapped |
|---|---|---|---|---|---|---|
| 313232 | Haemodialysis-associated hypotension | Observation | SNOMED | YES | YES | NO |
| 314432 | Maternal hypotension syndrome | Condition | SNOMED | YES | YES | NO |
| 317002 | Low blood pressure | Condition | SNOMED | NO | YES | NO |



**Negative controls**

For the negative control analyses, we used a subset of the 30 most prevalent negative control outcomes from the total of 76 that were considered in the reference study[1]: 1) human papilloma virus infection; 2) nicotine dependence; 3) verruca vulgaris; 4) acquired keratoderma; 5) Somatic dysfunction of lumbar region; 6) impacted cerumen; 7) acute conjunctivitis; 8) senile hyperkeratosis; 9) tobacco dependence syndrome; 10) sprain of ankle; 11) epidermoid cyst; 12) irregular periods; 13) genetic predisposition; 14) abnormal cervical smear; 15) wrist joint pain; 16) contact dermatitis; 17) tear film insufficiency; 18) onychomycosis due to dermatophyte; 19) injury of knee; 20) derangement of knee; 21) ingrowing nail; 22) chondromalacia of patella; 23) presbyopia; 24) leukorrhea; 25) herpes zoster without complication; 26) calcaneal spur; 27) ganglion cyst; 28) regular astigmatism; 29) acquired hallux valgus; 30) cervical stomatic dysfunction.

## DATA SOURCES

### CCAE

IBM MarketScan Commercial Claims and Encounters Database (CCAE) is a US employer-based private-payer administrative claims database. The data include adjudicated health insurance claims (e.g. inpatient, outpatient, and outpatient pharmacy) as well as enrolment data from large employers and health plans who provide private healthcare coverage to employees, their spouses, and dependents. Additionally, it captures laboratory tests for a subset of the covered lives. This administrative claims database includes a variety of free-for-service, preferred provider organisations and capitated health plans.

### MDCD

IBM MarketScan Multi-State Medicaid Database (MDCD) contains adjudicated US health insurance claims for Medicaid enrolees from multiple states and includes hospital discharge diagnoses and procedures, and outpatient pharmacy claims as well as ethnicity and Medicare eligibility. Members maintain the same identifier even if they leave the system for a brief period. The dataset lacks lab data.

### MDCR

IBM MarketScan Multi-State Medicare Supplemental and Coordination of Benefits Database (MDCR) represents health services of retirees in the US with primary or Medicare supplemental coverage through privately insured free-for-service, point-of-service, or capitated health plans. These data include adjudicated health insurance claims (e.g. inpatient, outpatient, and outpatient pharmacy). Additionally, it captures laboratory tests for a subset of the covered lives.



**ADDITIONAL RESULTS**

**Patient follow-up**

Table S1: Number of patients, person years, and events within quarters of predicted risk for acute myocardial infarction for the three main outcomes of the study in CCAE.

| Outcome | Risk quarter | ACE inhibitors | | | Beta blockers | | |
|---|---|---|---|---|---|---|---|
| | | Patients | Person years | Outcomes | Patients | Person years | Outcomes |
| acute myocardial infarction | 1 | 190,641 | 272,359 | 228 | 154,752 | 215,366 | 127 |
| | 2 | 242,256 | 355,184 | 631 | 103,137 | 151,554 | 318 |
| | 3 | 251,961 | 366,917 | 1,168 | 93,431 | 135,327 | 539 |
| | 4 | 239,338 | 333,513 | 2,075 | 106,055 | 146,365 | 1,508 |
| hospitalization with heart failure | 1 | 190,439 | 272,074 | 277 | 154,291 | 214,785 | 425 |
| | 2 | 241,930 | 354,725 | 579 | 102,734 | 150,991 | 427 |
| | 3 | 251,429 | 366,205 | 936 | 92,811 | 134,523 | 666 |
| | 4 | 237,868 | 331,593 | 1,905 | 103,453 | 143,091 | 1,857 |
| stroke | 1 | 190,200 | 271,769 | 357 | 154,164 | 214,606 | 356 |
| | 2 | 241,312 | 353,856 | 680 | 102,417 | 150,571 | 412 |
| | 3 | 250,160 | 364,467 | 1,005 | 92,309 | 133,831 | 531 |
| | 4 | 232,899 | 325,137 | 1,664 | 102,578 | 141,894 | 1,008 |



Table S2: Number of patients, person years, and events within quarters of predicted risk for acute myocardial infarction for the three main outcomes of the study in MDCD.

| Outcome | Risk quarter | ACE inhibitors | | | Beta blockers | | |
|---|---|---|---|---|---|---|---|
| | | Patients | Person years | Outcomes | Patients | Person years | Outcomes |
| acute myocardial infarction | 1 | 21,636 | 32,074 | 32 | 24,204 | 33,511 | 27 |
| | 2 | 28,974 | 42,897 | 151 | 16,863 | 24,769 | 95 |
| | 3 | 31,219 | 47,965 | 399 | 14,619 | 22,748 | 233 |
| | 4 | 25,217 | 39,654 | 866 | 20,621 | 31,740 | 1,006 |
| hospitalization with heart failure | 1 | 21,568 | 31,974 | 99 | 24,088 | 33,356 | 148 |
| | 2 | 28,820 | 42,665 | 327 | 16,645 | 24,460 | 249 |
| | 3 | 30,866 | 47,411 | 710 | 14,241 | 22,158 | 514 |
| | 4 | 23,484 | 36,969 | 1,596 | 17,288 | 26,967 | 1,857 |
| stroke | 1 | 21,512 | 31,891 | 66 | 24,058 | 33,286 | 89 |
| | 2 | 28,648 | 42,404 | 230 | 16,572 | 24,324 | 128 |
| | 3 | 30,456 | 46,764 | 501 | 14,112 | 21,979 | 316 |
| | 4 | 23,385 | 36,825 | 960 | 18,783 | 29,069 | 939 |



Table S3: Number of patients, person years, and events within quarters of predicted risk for acute myocardial infarction for the three main outcomes of the study in MDCR.

| Outcome | Risk quarter | ACE inhibitors | | | Beta blockers | | |
|---|---|---|---|---|---|---|---|
| | | Patients | Person years | Outcomes | Patients | Person years | Outcomes |
| acute myocardial infarction | 1 | 29,741 | 45,807 | 243 | 15,169 | 23,600 | 134 |
| | 2 | 29,087 | 44,749 | 360 | 15,822 | 24,743 | 218 |
| | 3 | 27,247 | 41,633 | 514 | 17,662 | 27,200 | 337 |
| | 4 | 20,830 | 31,071 | 647 | 24,080 | 35,277 | 791 |
| hospitalization with heart failure | 1 | 29,613 | 45,605 | 440 | 15,061 | 23,444 | 324 |
| | 2 | 28,911 | 44,487 | 585 | 15,622 | 24,456 | 477 |
| | 3 | 26,896 | 41,127 | 827 | 17,136 | 26,452 | 789 |
| | 4 | 19,846 | 29,708 | 1,070 | 21,853 | 32,397 | 1,704 |
| stroke | 1 | 29,190 | 45,008 | 378 | 14,878 | 23,171 | 245 |
| | 2 | 28,305 | 43,580 | 523 | 15,376 | 24,112 | 359 |
| | 3 | 26,009 | 39,815 | 680 | 16,832 | 26,011 | 544 |
| | 4 | 19,013 | 28,515 | 702 | 22,075 | 32,657 | 931 |



**Performance of prediction models**

Table S4: Discriminative ability (c-statistic) of the derived prediction models for hospitalization with heart failure in the matched set (development set), the treatment cohort, the comparator cohort, and the entire population in CCAE, MDCD, and MDCR.

| Cohort | CCAE | MDCD | MDCR |
|---|---|---|---|
| Matched | 0.77 (0.76, 0.78) | 0.81 (0.80, 0.81) | 0.74 (0.74, 0.75) |
| Treatment | 0.74 (0.73, 0.74) | 0.79 (0.78, 0.80) | 0.74 (0.73, 0.75) |
| Comparator | 0.77 (0.77, 0.78) | 0.82 (0.82, 0.83) | 0.73 (0.72, 0.74) |
| Entire population | 0.76 (0.75, 0.76) | 0.81 (0.80, 0.81) | 0.74 (0.74, 0.75) |



Table S5: Discriminative ability (c-statistic) of the derived prediction models for stroke in the matched set (development set), the treatment cohort, the comparator cohort, and the entire population in CCAE, MDCD, and MDCR.

| Cohort | CCAE | MDCD | MDCR |
|---|---|---|---|
| Matched | 0.71 (0.70, 0.72) | 0.77 (0.76, 0.78) | 0.66 (0.65, 0.67) |
| Treatment | 0.69 (0.68, 0.70) | 0.76 (0.75, 0.77) | 0.66 (0.65, 0.67) |
| Comparator | 0.71 (0.70, 0.72) | 0.79 (0.78, 0.80) | 0.65 (0.64, 0.67) |
| Entire population | 0.70 (0.69, 0.71) | 0.78 (0.77, 0.79) | 0.66 (0.65, 0.67) |



**Diagnostics**

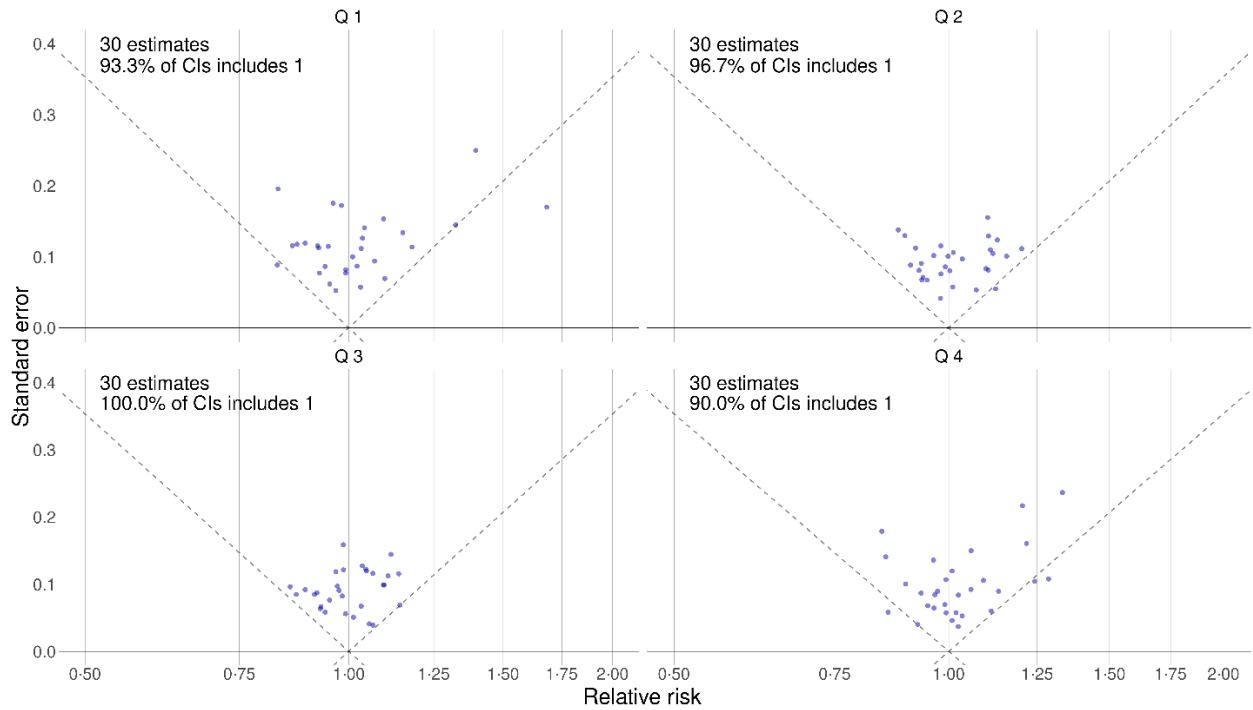

Figure S1: Systematic error in MDCD. Effect size estimates for the 30 negative controls (true hazard ratio = 1). Estimates below the diagonal dashed lines are statistically significant (alpha = 0.05) different from the true effect size. A well-calibrated estimator should have true effect size within the 95% confidence interval 95% of the times.



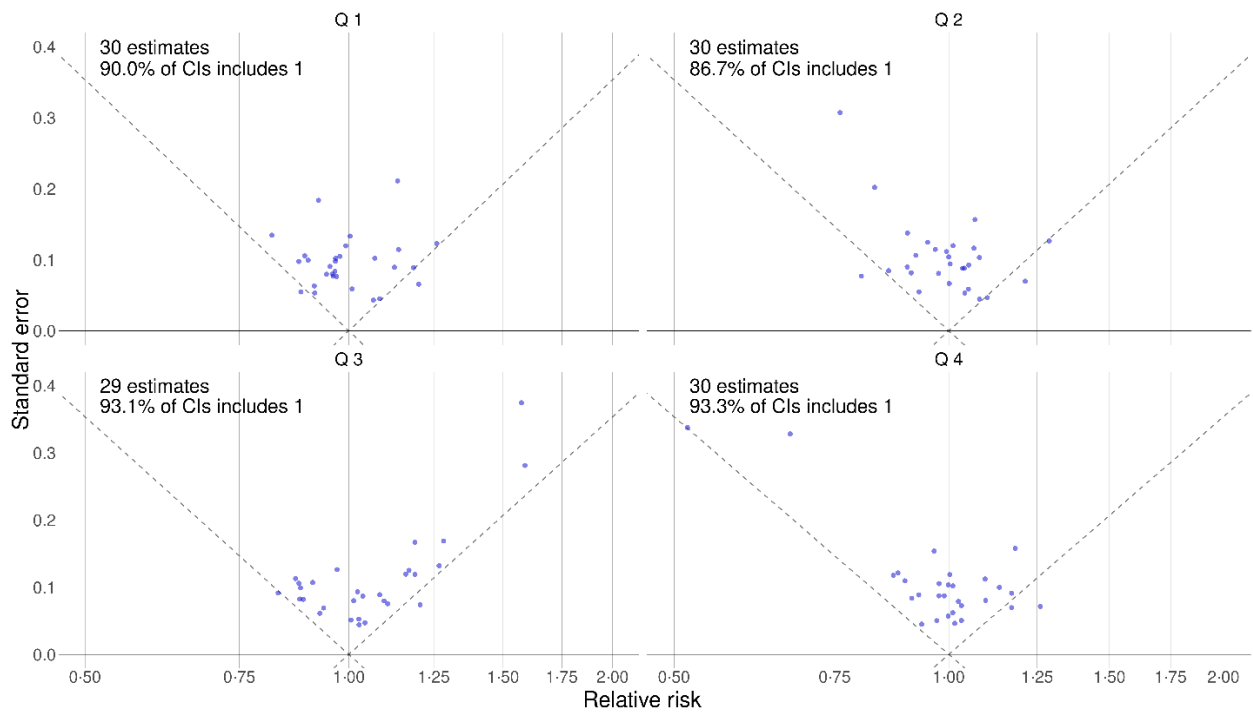

Figure S2: Systematic error in MDCR. Effect size estimates for the 30 negative controls (true hazard ratio = 1). In risk quarter Q3 one of the negative control effect estimates failed and therefore only the results for the remaining 29 are shown. Estimates below the diagonal dashed lines are statistically significant (alpha = 0.05) different from the true effect size. A well-calibrated estimator should have true effect size within the 95% confidence interval 95% of the times.



**Effect estimates within strata of hospitalization with heart failure and stroke**

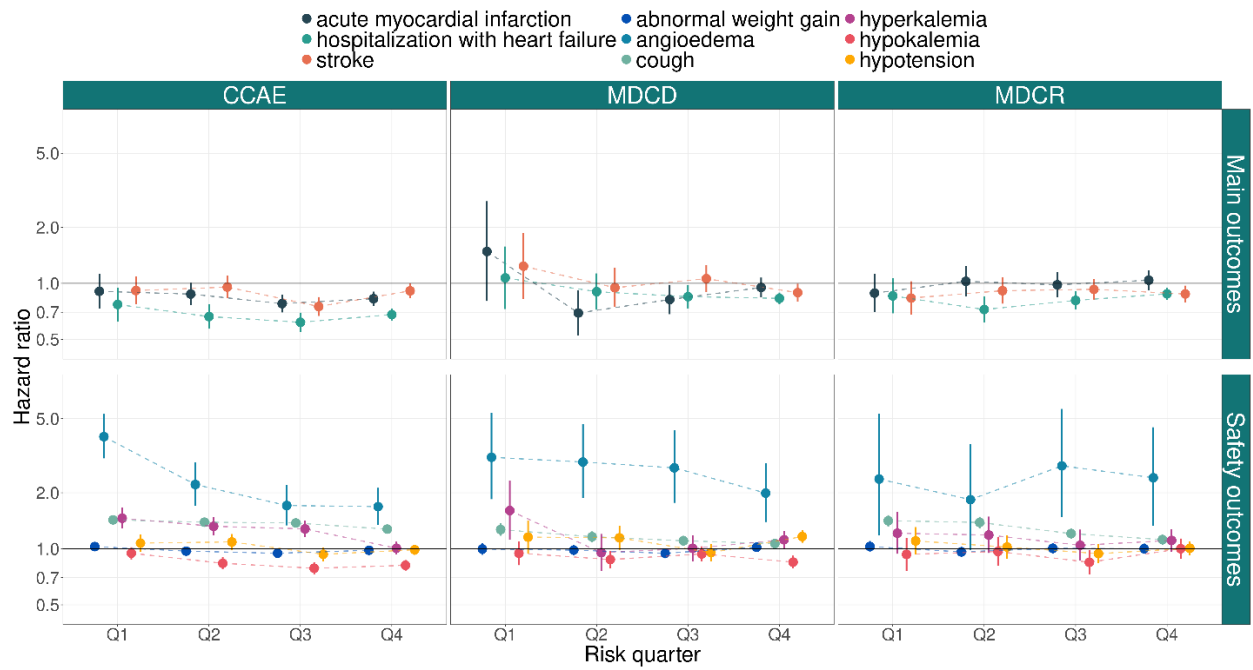

Figure S3: Overview of heterogeneity of ACE inhibitors treatment on the relative scale (hazard ratios) within strata of predicted risk of hospitalization with heart failure. Values below 1 favour ACE inhibitors, while values above 1 favour beta blockers.



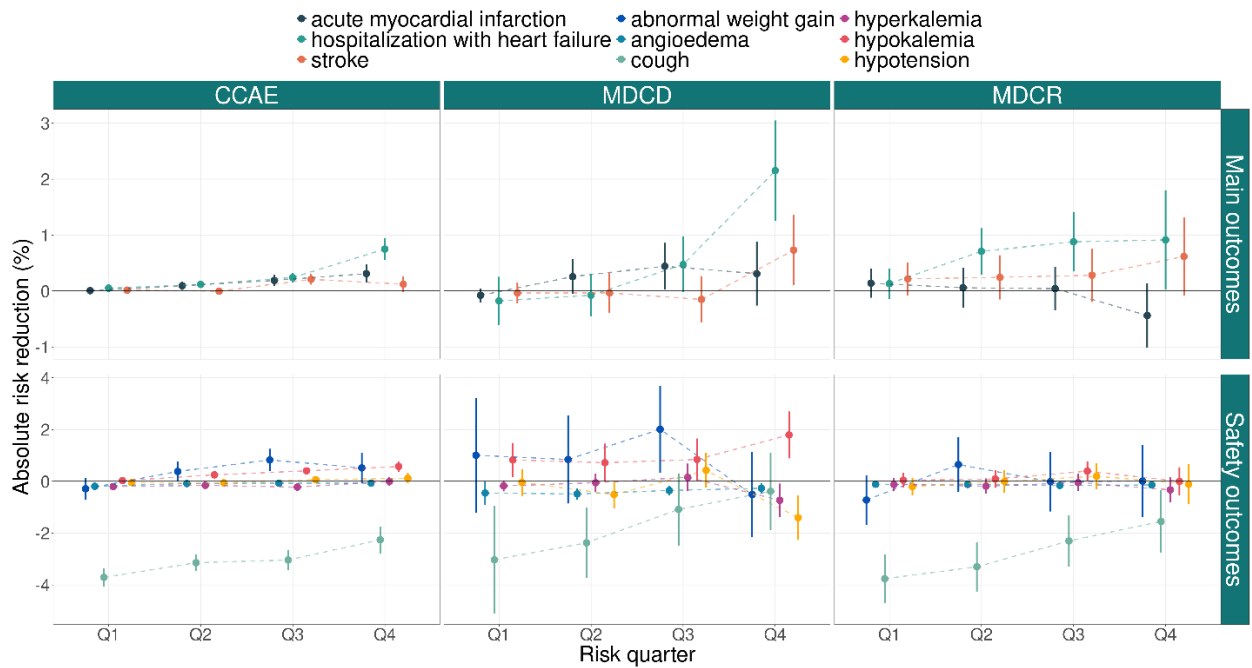

Figure S4: Overview of heterogeneity of ACE inhibitors on the absolute scale within strata of predicted risk of hospitalization with heart failure. Estimates of absolute treatment effect are derived as the difference in Kaplan-Meier estimates at 730 days after inclusion. Values above 0 favour ACE inhibitors, while values below 0 favour beta blockers.



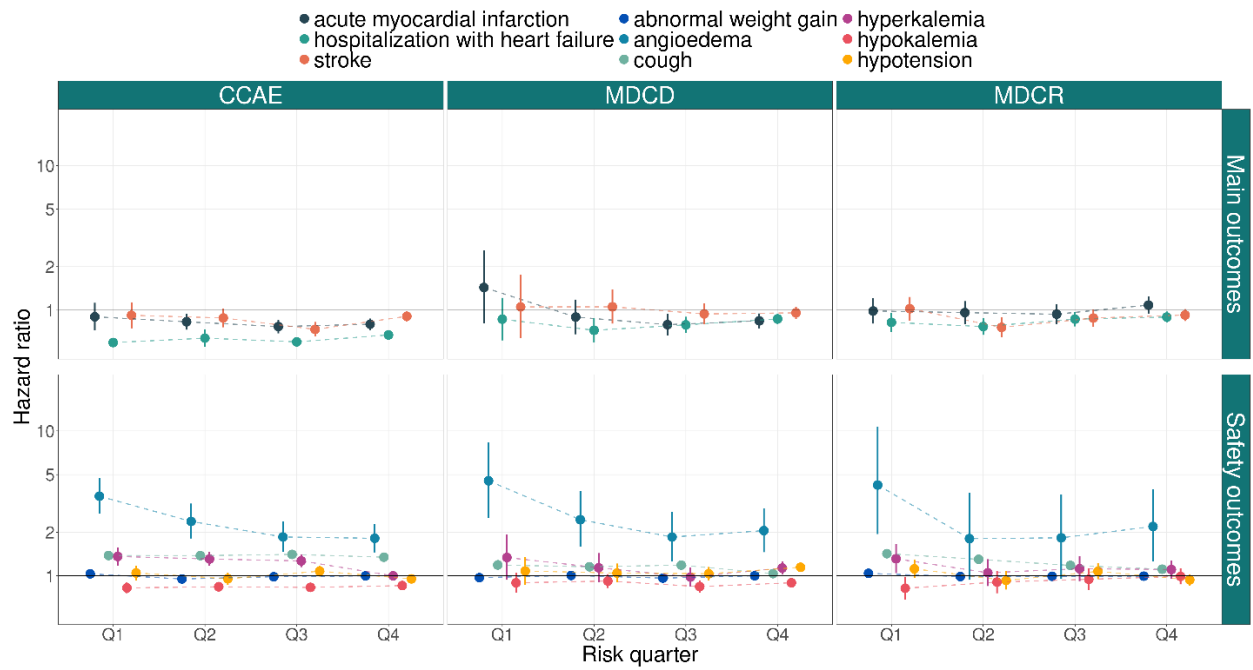

Figure S5: Overview of heterogeneity of ACE inhibitors treatment on the relative scale (hazard ratios) within strata of predicted risk of stroke (ischemic or hemorrhagic). Values below 1 favour ACE inhibitors, while values above 1 favour beta blockers.



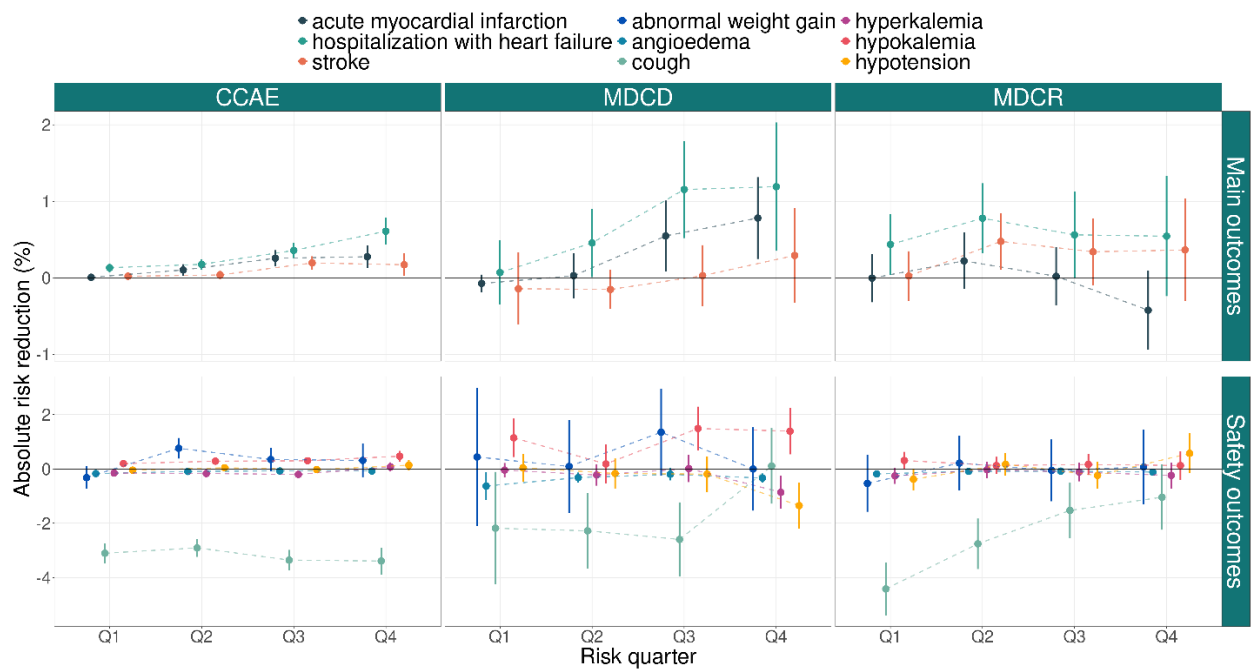

Figure S6: Overview of heterogeneity of ACE inhibitors on the absolute scale within strata of predicted risk of hospitalization with heart failure. Estimates of absolute treatment effect are derived as the difference in Kaplan-Meier estimates at 730 days after inclusion. Values above 0 favour ACE inhibitors, while values below 0 favour beta blockers.


\fi

\end{document}
\end{document}